\documentclass[10pt,twocolumn,letterpaper]{article}

\usepackage{iccv}
\usepackage{times}
\usepackage{epsfig}
\usepackage{amsmath}
\usepackage{amssymb}
\usepackage{multirow}
\usepackage{gensymb}
\usepackage{bm}
\usepackage{graphicx}
\usepackage{adjustbox}

\usepackage{float}
\usepackage[normalem]{ulem}
\useunder{\uline}{\ul}{}
\usepackage{subfigure}
\usepackage{algorithm}
\usepackage{algorithmic}
\renewcommand{\mathbf}{\boldsymbol}
\newcommand{\tx}{\bm{x}}

\newcommand{\tz}{\bm{z}}
\newcommand{\tv}{\bm{v}}

\newcommand{\ty}{\bm{y}}

\newcommand{\tR}{\bm{R}}

\newcommand{\tO}{\bm{O}}
\newcommand{\tn}{\bm{n}}

\newcommand{\talpha}{\bm{\alpha}}
\newcommand{\tbeta}{\bm{\beta}}
\newcommand{\tgamma}{\bm{\gamma}}

\DeclareMathOperator*{\argmin}{argmin}
\DeclareMathOperator*{\argmax}{argmax}

\usepackage[pagebackref=true,breaklinks=true,letterpaper=true,colorlinks,bookmarks=false]{hyperref}

\iccvfinalcopy 

\def\iccvPaperID{} 

\ificcvfinal\fi

\begin{document}

\title{Searching Efficient Model-guided Deep Network for Image Denoising}
\author{Qian Ning,\textsuperscript{\rm 1} ~ Weisheng Dong,\textsuperscript{\rm 1}\thanks{Corresponding author (wsdong@mail.xidian.edu.cn).} ~~ Xin Li,\textsuperscript{\rm 2} ~ Jinjian Wu,\textsuperscript{\rm 1} ~ Leida Li,\textsuperscript{\rm 1} ~ Guangming Shi\textsuperscript{\rm 1}\\ 
\textsuperscript{\rm 1}School of Artificial Intelligence, Xidian University ~~
\textsuperscript{\rm 2}West Virginia University \\ 
}

\maketitle
\ificcvfinal\thispagestyle{empty}\fi
\begin{abstract}
Neural architecture search (NAS) has recently reshaped our understanding on various vision tasks. Similar to the success of NAS in high-level vision tasks, it is possible to find a memory and computationally efficient solution via NAS with highly competent denoising performance. However, the {\em optimization gap} between the super-network and the sub-architectures has remained an open issue in both low-level and high-level vision. In this paper, we present a novel approach to filling in this gap by connecting model-guided design with NAS (MoD-NAS) and demonstrate its application into image denoising. Specifically, we propose to construct a new search space under model-guided framework and develop more stable and efficient differential search strategies. MoD-NAS employs a highly reusable width search strategy and a densely connected search block to automatically select the operations of each layer as well as network width and depth via gradient descent.  During the search process, the proposed MoG-NAS is capable of avoiding {\em mode collapse} due to the smoother search space designed under the model-guided framework. Experimental results on several popular datasets show that our MoD-NAS has achieved even better PSNR performance than current state-of-the-art methods with fewer parameters, lower number of flops, and less amount of testing time.    
\end{abstract}

\section{Introduction}
The field of image restoration, especially image denoising, has advanced rapidly in recent years. Many deep learning-based methods have achieved great performance in image denoising applications such as Trainable Nonlinear Reaction Diffusion Network (TNRD) \cite{chen:TPAMI:2017TNRD}, Denoising Convolutional Neural Network (DnCNN) \cite{zhang2017beyond}, Memory Network for Image Restoration (MemNet) \cite{tai2017memnet}, Non-local recurrent network (NLRN) \cite{liu2018non}, Evolutionary search for convolutional autoencoders (E-CAE) \cite{suganuma2018exploiting}, Dual residual networks (DuRN) \cite{liu2019dual}, and Neural nearest neighbors networks (N3Net) \cite{plotz2018neural}. 
Most recently, neural architecture search (NAS) based approach \cite{zhang2020memory} has been proposed and demonstrated highly competitive performance for the task of image denoising.   

Despite the extensive study of NAS in computer vision community, the main-stream applications of NAS have been limited to middle-to-high level vision tasks such as image classification \cite{zoph2017neural,liu2019darts}, semantic segmentation \cite{liu2019auto}, and object detection \cite{chen2019detnas,tan2019mnasnet}. Only a handful of works on leveraging NAS to low-level vision tasks have been published so far (e.g., image superresolution \cite{guo2020hierarchical,chu2019fast} and denoising  \cite{suganuma2018exploiting,zhang2020memory}). For example, hierarchical NAS has been studied for various low-level vision tasks such as image denoising \cite{zhang2020memory} and superresolution \cite{guo2020hierarchical}; however, they have adopted a similar search space to that in high-level vision tasks \cite{liu2019darts}. 
Similar to DARTS \cite{liu2019darts}, those NAS methods\cite{zhang2020memory,guo2020hierarchical} for low-level tasks have only searched the operations of layers, ignoring flexibility in terms of network width and depth.
Meanwhile, existing NAS strategies are known for suffering from the problem of instability, which is believed to arise from the notorious ``optimization gap'' between the super-network and its sub-architectures \cite{xie2020weight}.

The motivation behind this work is mainly two-fold. On one hand, inspired by recent works of model-guided network for image restoration \cite{dong:TPAMI:2018DPDNN,zhang2020deep}, we propose to construct a {\em new search space} by leveraging the domain knowledge implied in model-guided neural architecture.  To the best of our knowledge, this is the first work incorporating the domain knowledge of image restoration into NAS to design a new search space for low-level tasks. On the other hand, recent works on densely connected search space \cite{fang2020densely} and network pruning \cite{lu2020beyond} have motivated us to pursue more stable and flexible NAS strategies specifically tailored for the new constructed search spaces. Unlike previous NAS methods \cite{zhang2020memory,guo2020hierarchical,liu2019darts} that only searched the operations of each layer, our proposed method can automatically select not only the operations of each layer but also the network width and depth, achieving the objective of efficiency. 
The key contributions of this paper are summarized as follows.
\begin{itemize}
  \item We propose to search efficient model-guided deep networks for image denoising. Our approach consists of a new constructed search space and a tailored search strategy for selecting network width and depth automatically, achieving lightweight and low inferring time simultaneously. The new search space is built under model-guided framework, in which the deep denoiser is based on U-net. The construction of search space by model-guided design guarantees the stability of our method, avoiding mode collapse during the search process. 
  
  \item Selecting operations of each layer as well as the network width and depth is based on more flexible search strategies. By combining highly reusable width-search with densely connected blocks for depth search, efficient and effective networks have been searched for image denoising and compression artifact reduction.
  \item We have conducted different benchmark experiments on our searched network for image denoising and compression artifact reduction. Experimental results on several popular datasets show that our MoD-NAS performs comparably or even better than current state-of-the-art methods with fewer parameters, lower number of flops, and less amount of running time. 
\end{itemize}

\begin{figure*}[htbh]
\begin{center}
\includegraphics[width=1.0\linewidth]{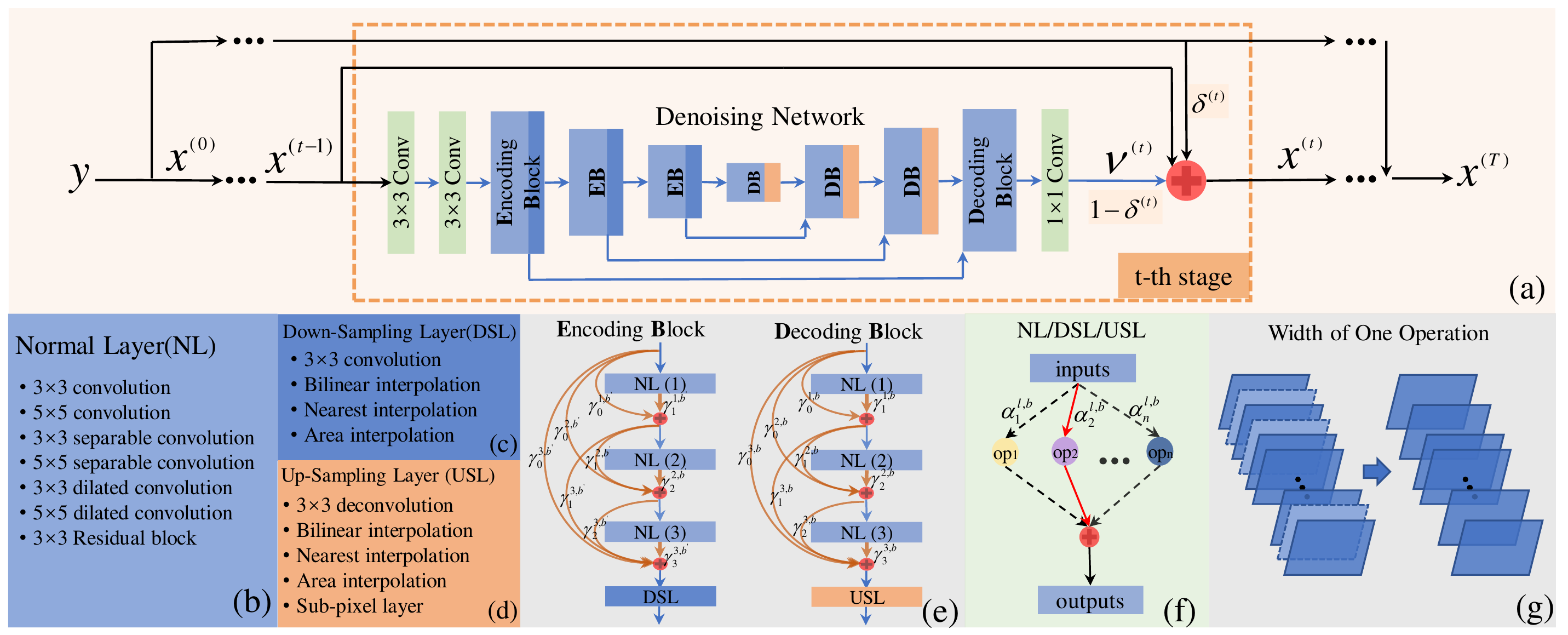}
\end{center}
\vspace{-5mm}
\caption{(a) The overall architecture of the proposed network; (b)-(d) a list of candidate operations to be searched for NL, DSL and USL respectively; (e) network depth searching; (f) layer operations searching; (g) network width searching.}
\vspace{-4mm}
\label{fig:framework}
\end{figure*}

\section{Related Works}
\subsection{Model-guided design for image denoising}
The problem of image denoising can be formulated as $\ty=\tx+\tn$, where $\tx,\ty$ denotes the clean/noisy image pair and $\tn$ denotes the additive white Gaussian noise. Based on the above model, we can obtain the clean image $\tx$ by Maximum a Posterior (MAP) estimation as
\begin{equation}
\footnotesize
\hat{\tx} = \argmax_{\tx} log P(\tx|\ty) = \argmax_{\tx} log P(\ty|\tx) +log P(\tx),\label{MoD}
\end{equation}
where $P(\ty|\tx)$ and $P(\tx)$ denote Gaussian likelihood and prior terms respectively, which can be expressed as

\begin{align}   
  P(\ty|\tx) &\propto exp(-\frac{1}{\sigma_n^2}||\ty-\tx||_2^2)           \\
  P(\tx) &\propto exp(-\lambda \tR(\tx)), \label{model-based_problem}
\end{align}
where $\sigma_n^2$ is the noise variance and $\tR(\tx)$ is the regularization function (e.g., sparsity-based \cite{kim:TPAMI:2010sparse}, nonlocal self-similarity based \cite{dong2013sparse,dabov:TIP:2007BM3D,zhang2017learning}). As mentioned in \cite{zhang2017learning,dong:TPAMI:2018DPDNN},  we can solve image denoising problem by leveraging a deep denoising network as a plug-and-play prior \cite{zhang2019deep}. Compared with hand-crafted priors, deep denoising priors achieve a more complex and flexible effect. With a mass of training data, deep network attempts to learn a nonlinear mapping function $\tv=f_{DN}(\tx)$ \cite{dong:TPAMI:2018DPDNN}, which can be used as a deep denoising prior as follows
\begin{equation}
P(\tx) \propto exp(-\lambda ||\tx-\tv||_2^2), \text{where} ~\tv = f_{DN}(\tx).
\end{equation}   
Therefore, Eq.~(\ref{MoD}) can be rewritten as 
\begin{equation}
\hat{\tx} = \argmin_{\tx} \frac{1}{\sigma_n^2}||\ty-\tx||_2^2 + \lambda ||\tx-\tv||_2^2, \label{model-based_problem_1}
\end{equation}
where $\tv = f_{DN}(\tx)$ and $\lambda$ denotes the regularization parameter.
Then, Eq. \eqref{model-based_problem_1} can  be solved by iterative regularization \cite{osher2005an} with a relaxation parameter $\delta^{(t)}=\frac{1}{1+\lambda^{(t)}\sigma_n^2}$,
\begin{subequations}
\begin{equation}\tv^{(t)} = f^{(t)}_{DN}(\tx^{(t-1)}) \label{solution_sub_problem_2} \end{equation} 
\begin{equation}\tx^{(t)} =  \frac{\ty+\lambda^{(t)}\sigma_n^2\tv^{(t)}}{1+\lambda^{(t)}\sigma_n^2} =\delta^{(t)}\ty+(1-\delta^{(t)})\tv^{(t)}. \label{solution_sub_problem_1} \end{equation}
\label{iter_equtation}
\end{subequations}

The basic idea of model-guided design (MoD) is to unfold conventional model-based iterative algorithms Eq.~\eqref{iter_equtation} into the implementation by cascaded DNN.  MoD-based IRCNN \cite{zhang2017learning} and DPDNN \cite{dong:TPAMI:2018DPDNN} have shown promising results on different image restoration tasks such as denoising, deblurring, and super-resolution. 
Since Eq. \eqref{solution_sub_problem_2} is solved by a manually designed denoising network, there is room for further improvement (e.g., via NAS) in terms of efficiency and better denoising performance. 

\subsection{Network architecture search (NAS)}
\noindent \textbf{Search Strategy.}
NAS has been proposed to overcome the difficulty of manually designing neural architectures for deep learning and achieved remarkable performance in various high-level tasks. Some early works have adopted reinforcement learning (RL) \cite{zoph2017neural,zoph2018learning} and evolutionary algorithm (EA)  \cite{liu2018hierarchical,real2019regularized}  as search strategies. However, both RL-based and EA-based methods require tremendous GPU resources and running time. For example, RL-based NASNet \cite{zoph2018learning} and EA-based AmoebaNet \cite{real2019regularized} take $48k$ and $76k$ GPU hours respectively. Given such prohibitive complexity, a differential search strategy such as DARTS \cite{liu2019darts} was proposed to relax the discrete search space by a differentiable proxy so NAS can be optimized by gradient descent.  Many recent works including ours and \cite{liu2019auto,wu2019fbnet,fang2020densely,zhang2020memory} have been inspired by this strategy of differential NAS. 

\noindent\textbf{Search Space Design.}
A cell-based search space has been proposed by NASNet \cite{zoph2018learning}, where the cell is defined by a directed acyclic graph with several nodes. Those cell-based methods \cite{zoph2018learning,real2019regularized,liu2019darts} search the operations between nodes (so-called feature maps) in a cell and repeat the cell to gain complete network architecture. However, networks found by cell-based search spaces often suffer from long inferring time. In order to improve on efficiency, a flurry of works such as ProxylessNAS \cite{cai2018proxylessnas}, FBNet \cite{wu2019fbnet}, DenseNAS \cite{fang2020densely} have proposed a new search space based on MobileNetV2 \cite{sandler2018mobilenetv2}. By searching for the expansion ratios and kernel sizes of MBConv layers, those NAS methods have often achieved better results than previous methods \cite{zoph2017neural,zoph2018learning,real2019regularized}. These works inspire us to construct a new search space based on another popular architecture (i.e., U-net) and automatically search the layer operations, network width and depth, achieving lightweight and low inferring time simultaneously. 

\noindent\textbf{NAS for Image Restoration.}
So far, there have been few works on applying NAS to image restoration. E-CAE \cite{suganuma2018exploiting} exploits for convolutional autoencoders for image inpainting and denoising by employing EA as search strategy, requiring enormous computational resource and costing a lot of time.  HiNAS \cite{zhang2020memory} employs gradient-based search strategy on cell-based search space for image denoising and deraining with less search time. Similar to HiNAS, we adopt gradient-based search strategy but design a new search space under model-guided framework for low-level tasks. Besides, our method can automatically search operations of each layer, network width and depth, while HiNAS has only searched the operations of each layer. 

\section{Model-guided Design with Neural Architecture Search (MoD-NAS)}
We first introduce proposed search space constructed under the model-guided framework. Then we present the strategies of searching operations for each layer as well as network width and depth. Finally, we summarize the overall search procedure of MoD-NAS.

\begin{figure*}[htbh]
\tiny
\newlength\structure
\setlength{\structure}{-5mm}
\centering
\begin{adjustbox}{valign=t}
\begin{tabular}{cccccc}
\hspace{-0.4cm}
\includegraphics[width=0.2\linewidth]{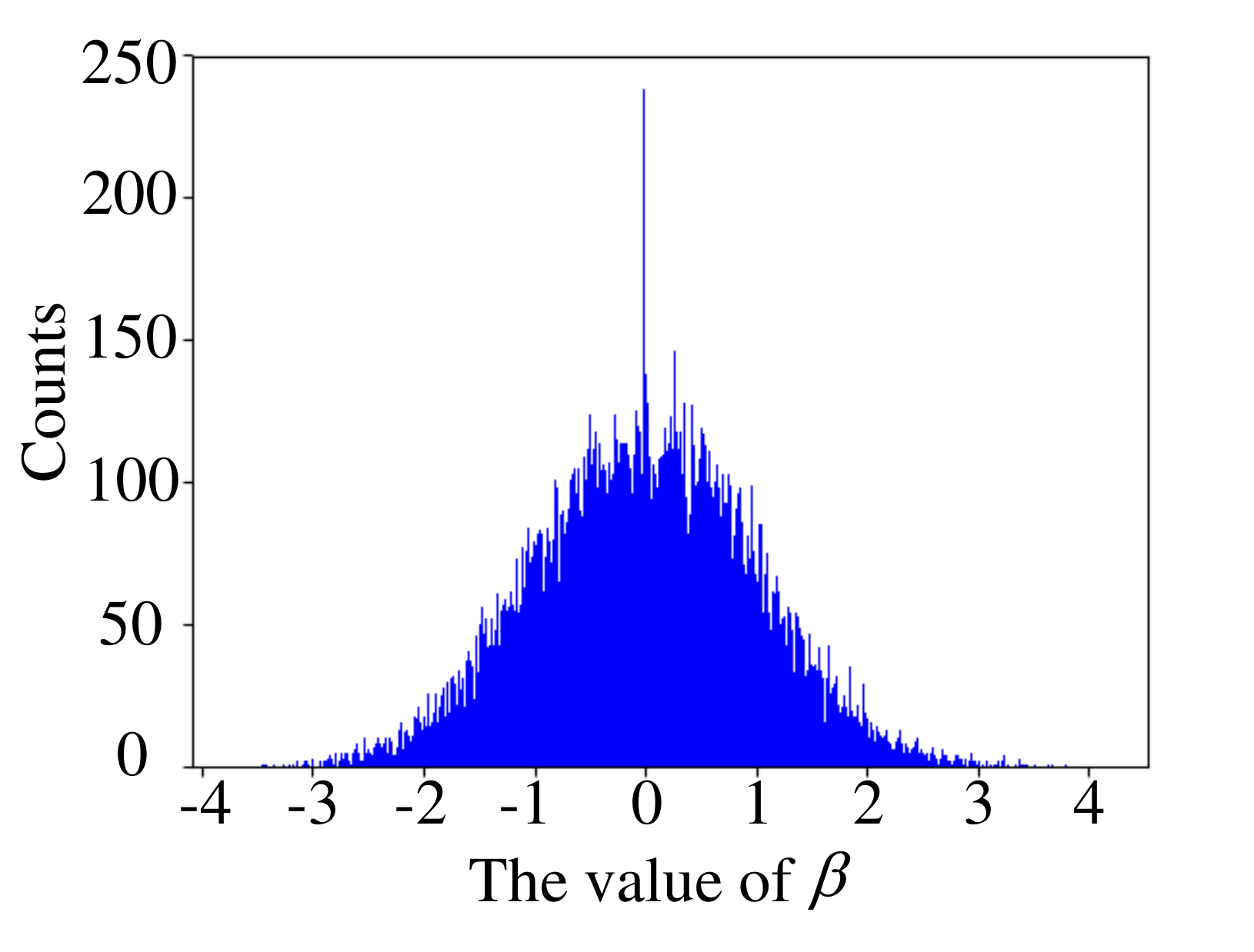} \hspace{\structure} &
\includegraphics[width=0.25\linewidth]{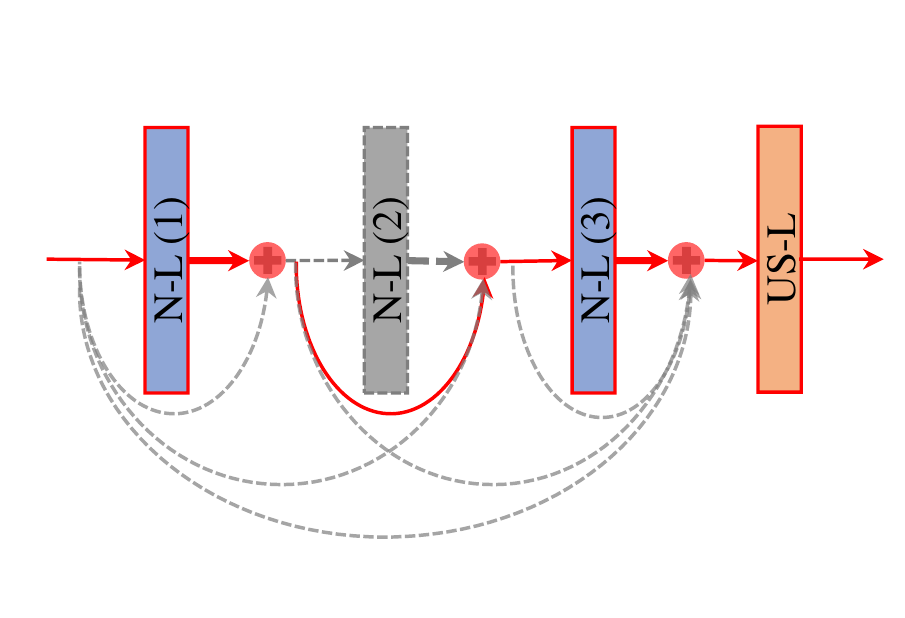} \hspace{\structure} &
\includegraphics[width=0.55\linewidth]{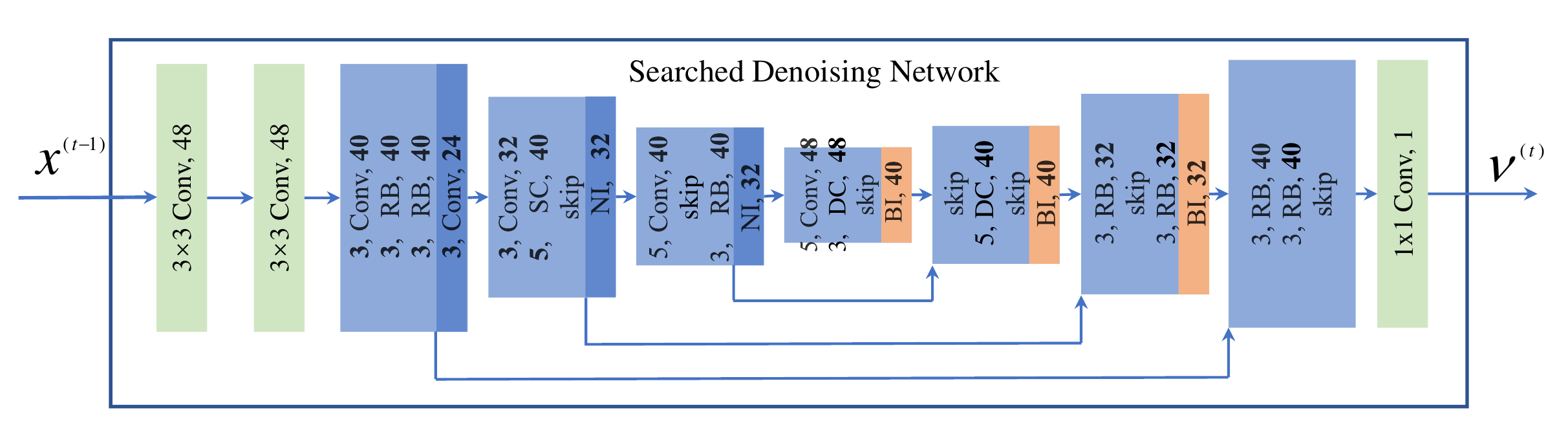} \hspace{\structure} & \\
 (a) \emph{The distribution of the width architecture parameters $\tbeta$} \hspace{\structure} &
 (b) \emph{An example of derived Decoding Block} \hspace{\structure} &
 (c) \emph{The searched network architecture} \hspace{\structure} &
 \end{tabular}
\end{adjustbox}
  \caption{ (a) The distribution of width architecture parameters $\tbeta$ (the x and y axes denote the values and counts of $\beta$ respectively); (b) an example of derived Decoding Block (the selected layer is highlighted by solid red bounding box and the discarded layer is shown by dashed gray bounding box); (c) the searched network architecture where RB denotes Residual Block, SC denotes Separable Convolution, DC denotes Dilated Convolution, NI denotes Nearest Interpolation, BI denotes Bilinear Interpolation. In each layer, the first number denotes the kernel size and the last number denotes the number of output channels (zoom in for better view). More details of searched networks can be found in appendix.}
  \vspace{-4mm}
  \label{fig:searched_network}
\end{figure*}

\subsection{The Search Space: Searchable U-net Under Model-guided Framework}
Mathematically, the solution to Eq. \eqref{model-based_problem_1} can be obtained by iteratively updating $\tv^{(t)}$ and $\tx^{(t)}$ following Eq.~(\ref{solution_sub_problem_2}) and Eq.~(\ref{solution_sub_problem_1}). By unfolding iterative updating algorithm through a deep network, model-guided methods \cite{zhang2017learning,dong:TPAMI:2018DPDNN,ning2020accurate} have achieved excellent performance. However, those networks are still hand-crafted and their optimality remains questionable. Under the model-guided framework, the key is to design a deep network $f_{DN}(\tx)$. As advocated in \cite{zhang2020memory}, NAS serves as an appealing remedy for optimizing neural architectures. Meanwhile, we focus on searching computationally efficient and lightweight denoising networks with additional domain knowledge of image denoising. 

Despite the U-net was proposed for medical image segmentation, the U-net has been demonstrated to have great performance in image restoration domain (e.g., dehazing~\cite{dong2020multi}, video deraining~\cite{yang2019frame}, video deblurring \cite{sim2019a}), showing the strong ability of denoising. Inspired by those success of U-net in image restoration, we propose to construct a new search space under model-guided framework, in which deep denoiser $f_{DN}(\tx)$ is based on U-net structure. When using U-net as a deep denoising prior, the one-step iteration Eq.~(\ref{iter_equtation}) can be unfolded into a deep network implementation as shown in Fig.~\ref{fig:framework}(a). The addition operator (red module) faithfully sums up the two terms in Eq.~\eqref{solution_sub_problem_1}. 

Note that the denoising network within the dashed orange box only represents one-step implementation after unfolding; its concatenation into multiple stages unfolds the iterative solution to Eq. \eqref{model-based_problem_1}.  Specifically, all layers of both encoding block (EB) and decoding block (DB) are searchable (highlighted by blue and orange modules). Taking DB as an example (refer to the right part of Fig.~\ref{fig:framework}(e)), except the last DB, each block consists of three normal layers and one upsampling layer. The last DB consists of three normal layers only.  Similarly, each EB consists of three normal layers and one down-sampling layer. During the search, there are $seven$, $four$ and $five$ candidate operations to be searched in normal layer, down-sampling layer and  up-sampling layer respectively. The detailed candidate operations for each layer are listed in Fig.~\ref{fig:framework}(b-d). The network width and depth can also be automatically selected via proposed searching strategies. By relaxing all discrete architectures of network space into differential formulations, we can search the architectures with gradient descent algorithms such as ADAM \cite{kingma2014adam}.

\subsection{The Search Strategies}
\noindent \textbf{Layer Operations}.
As shown in Fig.~\ref{fig:framework}(b-d), there are $seven$, $four$ and $five$ candidate operations to be searched in normal layer (NL), down-sampling layer (DSL) and  up-sampling layer (USL) respectively. It is worth mentioning that each convolution operation starts with a ReLU activation function but isn't followed by a batch normalization layer since it requires more GPU memory \cite{lim:CVPR:2017:EDSR}.  And each interpolation operation is followed by a $1\times1$ convolution for channel conversion. Taking the normal layer as an example, let $\mathbb{\tO}$ denotes the set of candidate operations listed in Fig.~\ref{fig:framework}(b). Every candidate operation $o\in\mathbb{\tO}$ of each layer $\ell$ in block $b$ has been allocated an architecture parameter $\alpha_o^{\ell,b}$. We have adopted the $softmax$ function to compute the architecture weight for every operation of layer $\ell$ in block $b$:
\begin{equation}
\omega_o^{\ell,b} =  \frac{exp(\alpha_o^{\ell,b})}{\sum _{o'\in\mathbb{\tO}}exp(\alpha_{o'}^{\ell,b})}.
\label{relax_operation}
\end{equation}
Finally, the output of operations in layer $\ell$ in block $b$ (refer to Fig.~\ref{fig:framework}(f)) can be expressed by
\begin{equation}
\tz^{\ell,b} =\sum _{o\in\mathbb{\tO}}\omega_o^{\ell,b}\cdot o(L^{\ell-1,b}),
\label{output_operation}
\end{equation}
where $L^{\ell-1,b}$ denotes the inputs of layer $\ell$ in block $b$. In a similar manner, all other types of layers can also be relaxed.

In summary, the task of choosing the best operation for layer $\ell$ in block $b$ has been translated to the problem of optimizing architecture parameters $\alpha^{\ell,b}$ which can be solved through gradient descent algorithms \cite{liu2019darts}. After the supernet is trained, we only need to choose the operation $\hat{o}^{\ell,b}$ with the largest architecture weight $\omega^{\ell,b}_{\hat{o}}$ computed by $\alpha^{\ell,b}$ with Eq.~\eqref{relax_operation} and discard the others. In other words, the selection of layer operations can be formulated into the following:
\begin{equation}
\hat{o}^{\ell,b}= \argmax_{o\in\mathbb{\tO}}  \omega_o^{\ell,b} .
\label{selection_operation}
\end{equation}
As shown in Fig.~\ref{fig:framework}(f), only one operational pathway highlighted by solid red arrows has been selected. 

\noindent \textbf{Network Width}.
HiNAS \cite{zhang2020memory} and  Autolab \cite{liu2019auto} search the width of a cell by stacking cells with different widths side-by-side (i.e., $1W$, $2W$, and $4W$, where $W$ is the basic width which has to be set manually before the search). Such strategy suffers from prohibitive cost of computation, GPU memory and searching time. Besides, this search strategy can only search a finite number of width parameters (as determined by $W$) and all layers in one cell have to share the same width. It follows that many potential architectures (e.g., with varying layer width in one cell) are excluded from the search, implying the lack of flexibility.

Inspired by the rapid advances in network pruning \cite{lu2020beyond}, we propose a new {\em highly reusable width-search} method for searching the width of every layer. As shown in the left part of Fig.~\ref{fig:framework}(g), every channel of operation $o$ of layer $\ell$ in block $b$ has been assigned an architecture parameter $\beta_c^{o,\ell,b}$, where $c$ denotes the \emph{c-th} channel of totally $C$ channels.  During the search process, the width and other architecture parameters will be optimized together by the gradient descent algorithm. 
Once the supernet has been trained, the probability distribution of the width architecture parameters $\tbeta$ typically observes a heavy-tailed distribution with a single peak at the origin (as shown in Fig.~\ref{fig:searched_network}(a)). Such observation implies that we can discard the channel with small architecture parameters $\beta_c^{o,\ell,b}$ around zero. Specifically, we have chosen the top-$\hat{M}$ channels with the largest architecture parameters $\beta_c^{o,\ell,b}$. The criterion for our selection can be written as:
\begin{subequations}
\begin{equation}\sum_{i=1}^{\hat{M}}|\beta_i^{o,\ell,b}|\geq 90\% \cdot {\sum_{j=1}^{C}|\beta_j^{o,\ell,b}|} \label{proportion} \end{equation}
\begin{equation}\hat{M} \mod 2^n = 0,\label{gpu_acceleration} \end{equation}
\label{selection_width}
\end{subequations}
where Eq. \eqref{proportion} reflects the idea of preserving large parameters only and Eq. \eqref{gpu_acceleration} is used for GPU acceleration (we have set $n=3$ in our experiment) \cite{fang2020densely}.
After the pruning, preserved channels are shown in the right part of Fig.~\ref{fig:framework}(g). Note that the actual number of preserved channels is a variable, which makes our architecture more flexible than fixed-width searching such as HiNAS \cite{zhang2020memory} and E-CAE \cite{suganuma2018exploiting}.

\noindent \textbf{Network Depth}.
Searching the suitable depth of a network is crucial to the success of NAS-related applications. Previous works \cite{xie2020weight,wu2019fbnet,cai2018proxylessnas} usually search the depth of network by adding a skip operation in the candidates. Once a skip operation has been selected, it is equivalent to the reduction of network depth by one layer. An undesirable consequence of such search strategy is that the produced network might be too shallow when many skip operations have been selected. One possible explanation of this phenomenon is that the skip has the same probability as the other operations and skip is often more frequently selected during search because it does not have any parameter. In this paper, we propose a new {\em densely connected block} to address this issue by utilizing dense connections \cite{zhang2018residual} to implement the strategy of searching the suitable depth of the network. 

As shown in Fig.~\ref{fig:framework}(e), every candidate connection highlighted by orange color have been assigned an architecture parameter. Namely the path from layer $i$ to layer $\ell$ in block $b$ has a parameter $\gamma_i^{\ell,b}$. Similar to the relaxation of layer operations, we adopt $softmax$ function to compute the probability $p_{i}^{\ell,b}$ of each path.
 Let $L^{\ell,b}$ denotes the outputs of layer $\ell$ of block $b$, then
\begin{equation}
\begin{split}
&L^{\ell,b} =\sum _{i=0}^{\ell-1} p_{i}^{\ell,b}\cdot L^{i,b}+p_{\ell}^{\ell,b}\cdot\tz^{\ell,b}  \\
&p_{i}^{\ell,b} =  \frac{exp(\gamma_i^{\ell,b})}{\sum _{j=0}^{\ell}exp(\gamma_{j}^{\ell,b})}, \label{selection_depth}
\end{split}
\end{equation}
where $\tz^{\ell,b}$ denotes the output of operations in layer $\ell$ in block $b$. After the supernet is trained, we only choose the paths with the largest probability $p_{i}^{\ell,b}$ in each layer and discard the others as shown in
 Fig.~\ref{fig:searched_network}(b). 


\subsection{Overall Search Procedure}
\begin{algorithm}[tbh]
\caption{Proposed MoD-NAS Algorithm.}
$\bullet$ \textbf{Initialization}:

    \hspace{0.5cm} (1) Initialize $\tx$ as $\tx^{(0)}=\ty$.
    
$\bullet$ \textbf{While} the search process not converge \textbf{do}

    \hspace{0.5cm} (1) \textbf{Forward inferring}: for $t=1,2,......,T$, \textbf{do} 
    
    \hspace{0.8cm} a) Compute $\tv^{(t)}=f_{DN}^{(t)}(x^{(t-1)})$;
    
    \hspace{0.8cm} b) Compute $\tx^{(t)}$ via Eq. \eqref{solution_sub_problem_1};
    
    \hspace{0.8cm} c) $t=t+1$.
    
  \hspace{0.8cm} \textbf{End for}
   
  \hspace{0.5cm} (2) \textbf{Backpropagation}:
   
  \hspace{0.8cm} a) Update weights by descending $\nabla\mathbb{W}\mathcal{L}_{train}(\mathbb{W},\mathbb{A})$;
 
  \hspace{0.8cm} b) Update architecture parameters by descending
   
  \hspace{1.1cm} $\nabla\mathbb{A}\mathcal{L}_{val}(\mathbb{W},\mathbb{A})$.
 
$\bullet$ \textbf{Derive the final architecture}: 




 \hspace{0.5cm} (1) derive the final architecture based on the depth parameter $\tgamma$, operations parameters $\talpha$ via Eq. \eqref{selection_operation} and  width parameters $\tbeta$ via Eq. \eqref{selection_width} in sequence.

\label{Algorithm}
\end{algorithm}

Putting things together, we can see how the search space constructed by MoD can be seamlessly integrated with the strategy of NAS as follows (refer to Algorithm \ref{Algorithm}). The unfolding process can be summarized into the procedure of \textbf{Forward inferring} of Algorithm \ref{Algorithm}. During this procedure, the mapping $f_{DN}^{(t)}$ corresponds to the denoising network in Fig.~\ref{fig:framework}(a) at \emph{t-th} stage. 
Our MoD-NAS adopts a differentiable search strategy, where the search process can be optimized by gradient descent algorithms such as ADAM \cite{kingma2014adam}. 
We train the supernet with the following MSE loss:
\begin{equation}
(\mathbb{W},\mathbb{A}) = \argmin_{\mathbb{W},\mathbb{A}}\sum^N_{i=1}\|\mathcal{F}(\ty_i;\mathbb{W},\mathbb{A})-\tx_i\|_2^2,
\label{loss_function_search}
\end{equation}
where $\ty_i$ and $\tx_i$ denote the $i$-th pair of degraded and original image patches respectively and $\mathcal{F}(\ty_i;\mathbb{W},\mathbb{A})$ denotes the reconstructed image patch by the supernet with the parameters set of operation weights $\mathbb{W}$ and the parameters set of architecture $\mathbb{A}$. 
We alternatively optimize the operation weights by descending $\nabla\mathbb{W}\mathcal{L}_{train}(\mathbb{W},\mathbb{A})$ on the training set , and optimize the architecture parameters by descending $\nabla\mathbb{A}\mathcal{L}_{val}(\mathbb{W},\mathbb{A})$ on the validation set as shown in procedure \textbf{Backpropagation} of Algorithm \ref{Algorithm}. 
When the supernet has been trained, we derive the final architecture based on the parameters $\talpha,\tbeta,\tgamma$ as shown in Algorithm \ref{Algorithm}. An example of searching for the final denoising network architecture is shown in Fig.~\ref{fig:searched_network}(c).
\vspace{-3mm}

\section{Experimental Results}
\vspace{-1mm}
\subsection{Experimental Settings}
\noindent\textbf{Benchmark Datasets.} We have randomly selected 4000 images from the Waterloo dataset for training. Following  \cite{zhang2017beyond,plotz2018neural,dong:TPAMI:2018DPDNN,jia2019focnet}, three standard benchmark datasets (Set12, BSD68, Urban100) are used for testing. The noisy images are generated by adding white Gaussian noise to the corresponding clean images with $\sigma=15,25,50$ following \cite{zhang2017beyond,dong:TPAMI:2018DPDNN,jia2019focnet}. It is worth mentioning that the search experiments are conducted with $\sigma=25$ and the training experiments are conducted with $\sigma=15,25,50$. 

\begin{figure}[htbh]
\begin{center}
\includegraphics[height=2.0in]{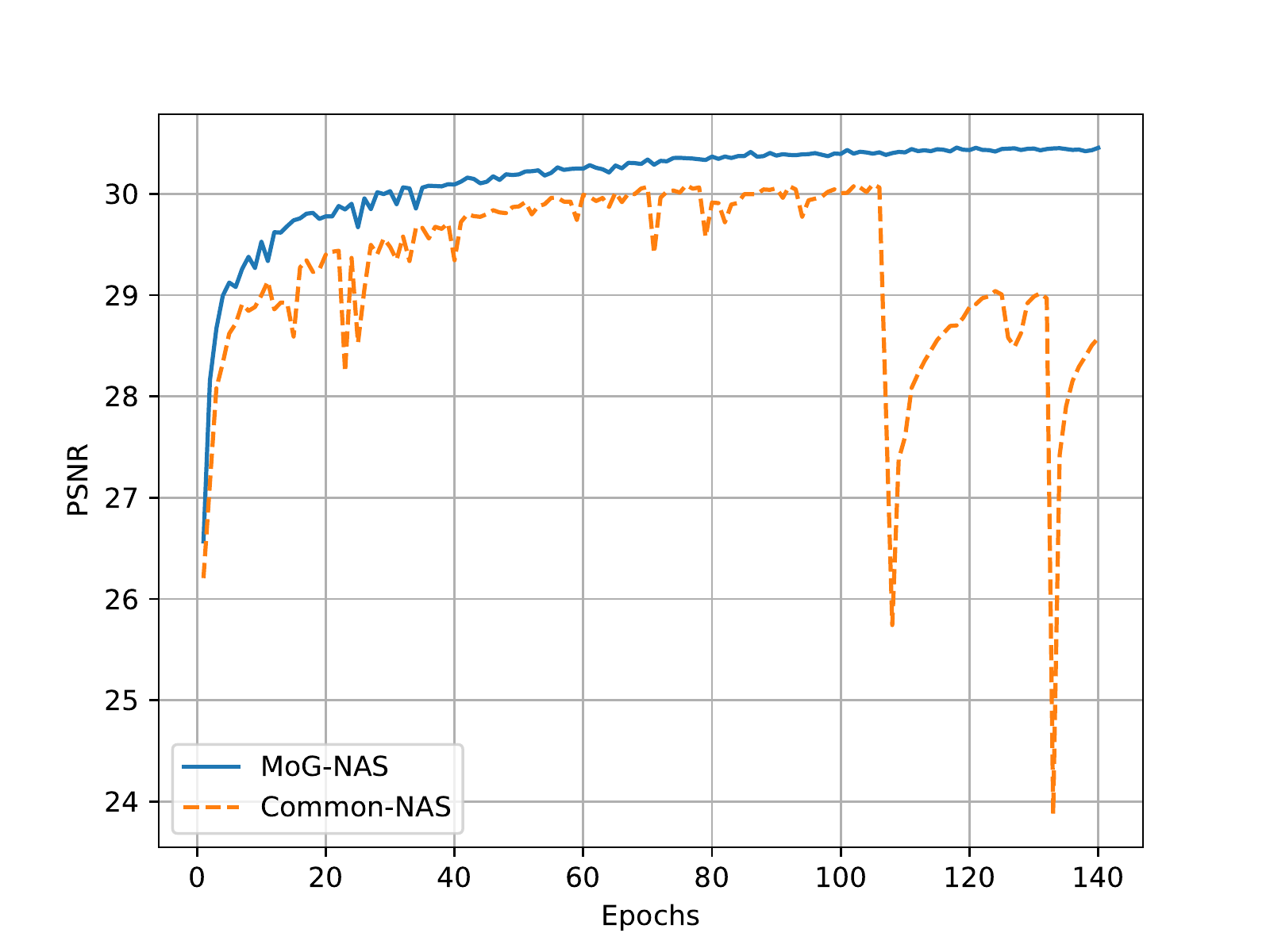}
\end{center}
\vspace{-0.2in}
\caption{The performance of supernet evaluated on Set12 dataset during search process with $\sigma=25$. The blue line refers to the search process with MoD-NAS and orange line refers to the search process with differentiable NAS (without MoD-NAS). }
\vspace{-0.2in}
\label{fig:benifit_MoD-NAS}
\end{figure}

\noindent\textbf{Search Settings.}
The training dataset \cite{ma2017waterloo} has been equally divided into nonoverlapping two parts: one for updating the weights of network operations (Training $\mathbb{W}$) and the other for updating the architecture parameters (Training $\mathbb{A}$).  We randomly select 12 noisy patches sized by $64\times64$ as the inputs. Two ADAM optimizers  \cite{kingma2014adam} with $\beta_1=0.9, \beta_2=0.999, \epsilon=10^{-8}$ are adopted to optimize parameters sets $\mathbb{W}$ and $\mathbb{A}$ respectively. The learning rate of two optimizers decays from $10^{-3}$ to $10^{-5}$ with the cosine annealing schedule \cite{loshchilov2016sgdr} within $140$ epochs. The stages of supernet is set to $T=2$ and the initial number of channels is set to $C=48$. In the first 40 epochs, we only update the parameters set $\mathbb{W}$ of operations and the parameters sets $(\mathbb{W},\mathbb{A})$ are optimized alternately in the remaining epochs. Our network is implemented by the Pytorch framework and the total searching time takes about \textbf{7} hours using one NVIDIA 2080Ti GPU.

\noindent\textbf{Training Settings.}
We train the network that searched by our method with MSE loss. 
We randomly select 32 noisy patches sized by $128\times128$ as the inputs. The ADAM algorithm  \cite{kingma2014adam} with $\beta_1=0.9, \beta_2=0.999, \epsilon=10^{-8}$ is adopted to optimize the network. The learning rate decays from $10^{-3}$ to $10^{-5}$ with the cosine annealing schedule \cite{loshchilov2016sgdr} within $600$ epochs. The searched denoising network is shown in Fig.~\ref{fig:searched_network}(c). Our network is implemented by  Pytorch and the total training takes about \textbf{10} hours when $T=3$ and using one NVIDIA RTX 2080Ti GPU.
\vspace{-0.2in}
\begin{figure}[htbh]
\begin{center}
\includegraphics[height=2.0in]{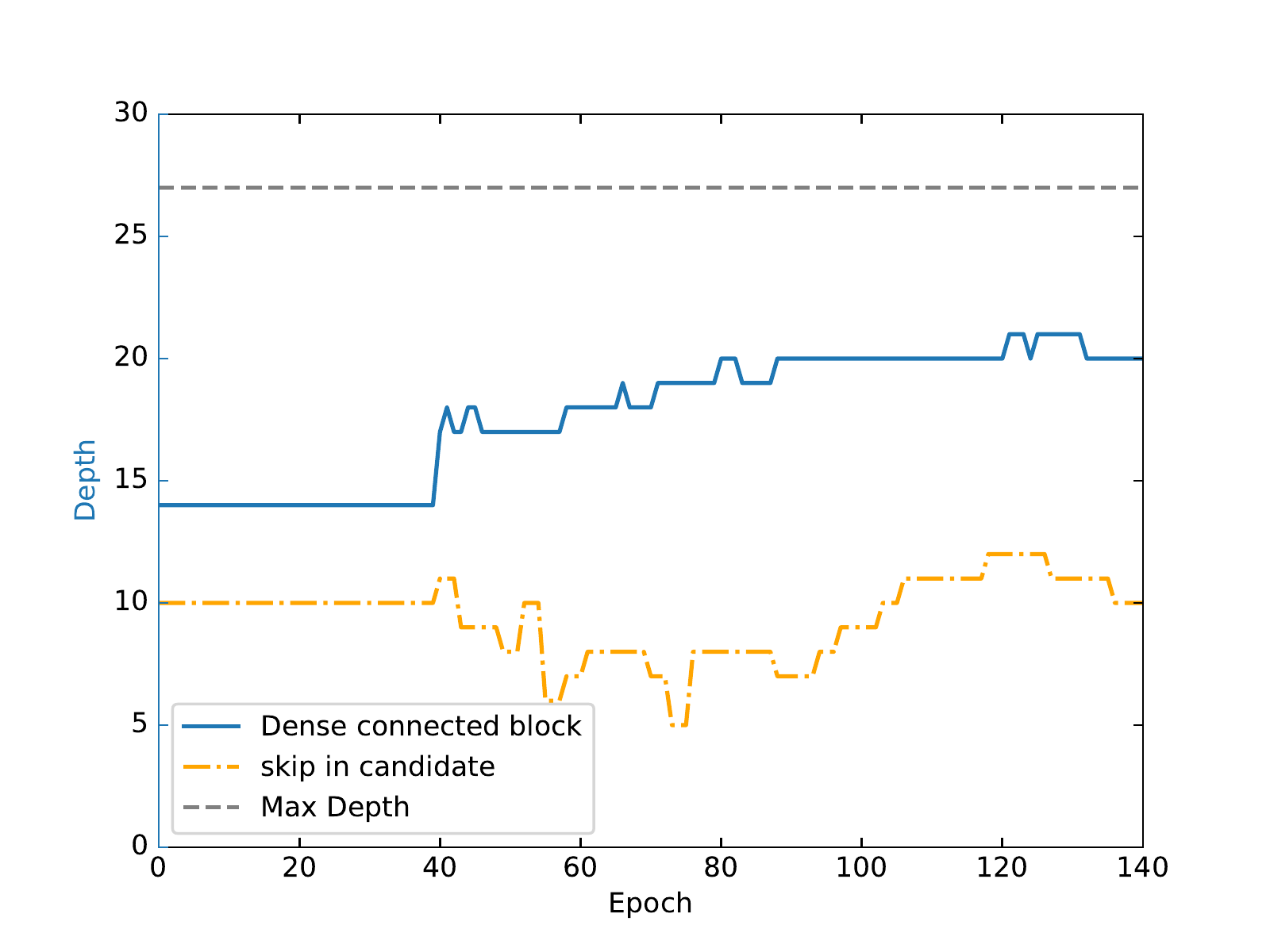}
\end{center}
\vspace{-0.2in}
\caption{The benefit of searching for network depth. The blue line refers to the search process with dense connected block and orange line refers to the search process adding skip in candidates and gray line refers to the max depth of one U-net. }
\vspace{-0.2in}
\label{fig:benifit_depth}
\end{figure}

\subsection{Ablation Study}
\noindent \textbf{Benifits of MoD-NAS.}
During the searching process of differentiable NAS,  there is a phenomenon that the performance of supernet drops a lot when the number of search epochs becomes large. This phenomenon known as {\em mode collapse} has been observed in both high-level tasks\cite{liang2020darts+} and low-level tasks \cite{zhang2020memory}. DATRS+ \cite{liang2020darts+} for image classification and HiNAS \cite{zhang2020memory} for image denoising have employed a similar way called \emph{early stopping}. However, this heuristic strategy does not solve the problem of mode collapse but gets around it. 

In this paper, by incorporating model-guided design with differentiable NAS (MoD-NAS), the performance of supernet remains stable and even increases slightly when the number of search epochs becomes large. To demonstrate this advantage, we have compared the performance of our MoD-NAS against common differentiable NAS \cite{liu2019darts} during the search process evaluated on the Set12 dataset as shown in Fig.~\ref{fig:benifit_MoD-NAS}. Note that we have removed the model-guided design part by only searching one denoising network as the baseline control; while all other settings are kept the same as MoD-NAS. From Fig.~\ref{fig:benifit_MoD-NAS}, we can see that the performance of proposed MoD-NAS stays stable during the whole search process, while the baseline method suffers from apparent collapses, especially when $Epochs>100$. Besides, during the search process, our proposed MoD-NAS demonstrates improved stability, showing a much smoother PSNR curve than the baseline. 
We argue that such benefit of the proposed MoD-NAS method can be explained away by MoD-NAS providing a smoother search space than differentiable NAS \cite{liu2019darts}.
Therefore, the proposed MoD-NAS enjoys excellent convergence property thanks to its search space is built under mode-guided framework with domain knowledge.

\noindent \textbf{Benefits of Searching for Network Width.}
To verify the effectiveness of searching network width, we have conducted the experiment MoD-NAS(T=3) ($C=64$), which changes all channels of MoD-NAS(T=3) to $64$. The experiment results of different widths of MoD-NAS(T=3) have been shown in Tab.~\ref{tab:abl_width}, from which we can see that MoD-NAS(T=3) achieved almost the same result on Set12 dataset. Therefore, MoD-NAS(T=3) achieves a better trade-off between the number of parameters or flops and accuracy in searching for network width, reaching the goal of efficiency.

\begin{table}[htbh]
\centering{
\caption{Comparisons of different width of MoD-NAS(T=3) on Set12 dataset with $\sigma=25$.
 }
\label{tab:abl_width}
\resizebox{1.0\columnwidth}{!}{
\begin{tabular}{|c|c|c|c|}
\hline
Models & parameters &flops  &PSNR\\ \hline
  MoD-NAS(T=3) &1256k  &1.45G &30.88 \\
  MoD-NAS(T=3) ($C=64$) &3245k  &3.46G &30.89 \\ \hline

\end{tabular}}}
\end{table}

\begin{table*}[htbh]
\centering{
\caption{Average PSNR results  for \textbf{Gaussian} image denoising on three benchmark datasets.
 The best performance is shown by bold and the second best performance is shown by underline. The testing time is the total time of evaluating the whole Urban100 dataset.}
\label{tab:guassian_denoising}
\resizebox{2.0\columnwidth}{!}{
\begin{minipage}{1.0\textwidth}
\begin{tabular}{|c|c|c|c|c|c|c|c|c|c|c|c|c|}
\hline
\multirow{2}{*}{Methods} & \multicolumn{3}{c|}{Set12}                      & \multicolumn{3}{c|}{BSD68}                       & \multicolumn{3}{c|}{Urban100}                    & \multirow{2}{*}{params} & \multirow{2}{*}{flops} & \multirow{2}{*}{{\begin{tabular}[c]{@{}c@{}}testing \\ time\end{tabular}}} \\ \cline{2-10}
                        & 15             & 25             & 50            & 15             & 25             & 50             & 15             & 25             & 50             &                         &                        &                        \\ \hline \hline
BM3D \cite{dabov:TIP:2007BM3D}                   & 32.37          & 29.97          & 26.72         & 31.07          & 28.57          & 25.62          & 32.35          & 29.70           & 25.95          & -                       & -                      &-                        \\ 
TNRD \cite{chen:TPAMI:2017TNRD}                    & 32.50           & 30.06          & 26.81         & 31.42          & 28.92          & 25.97          & 31.86          & 29.25          & 25.88          & -                        &  -                      &-                        \\
DnCNN \cite{zhang2017beyond}                  & 32.86          & 30.44          & 27.18         & 31.73          & 29.23          & 26.23          & 32.68          & 29.97          & 26.28          & {\ul 556k}                    & {\ul 1.28G}                  &21.45s                        \\ 
N3Net \cite{plotz2018neural}                  & -              & 30.55          & 27.43         & -              & 29.30           & 26.39          & -              & 30.19          & 26.82          & 706k                    & 1.62G                  &594.38s                        \\
MemNet \cite{tai2017memnet}                 & 32.96          & 30.60           & 27.03         & 30.76          & 29.19          & 25.25          & 32.33          & 30.68          & 25.56          & 2905k                   & 6.69G                  & 93.83s                       \\ 
DPDNN  \cite{dong:TPAMI:2018DPDNN} \footnote{DPDNN can be viewed as a baseline method since it is a model-guided method and the denoising network is manually designed based on U-net. }               & 32.91          & 30.54          & 27.50          & 31.83          & 29.27          & 26.40           & 32.98          & 30.30          & 26.85          & 1363k                   & 5.25G                  &68.62s                        \\ 
FOCNet  \cite{jia2019focnet}                 & 33.07          & 30.73          & 27.68          & 31.83          & 29.38          & 26.50           & 33.15          & 30.64          & {\ul 27.40 }         & -                   & -                  &125.13s                        \\ 
RDN \cite{zhang2020residual}                    & 32.95          & 30.66          & 27.60          & 31.74          & 29.29          & 26.41          & 33.04          & 30.50          & {\ul 27.40} & 21970k                  & 50.6G                  &386.95s                        \\ \hline
E-CAE \cite{suganuma2018exploiting}                & -               & -               & 26.53           & -               & -               & 25.86           & -               & -               & 24.51           & 1062k         & 2.45G          & 40.04s \\
HiNAS  \cite{zhang2020memory}  \footnote{Since the code of HiNAS is not publicly available, we have tried our best to reproduce HiNAS based on the technical details of the original paper. }              & 32.50           & 30.35           & 27.25           & 31.16           & 28.92           & 26.04           & 31.92           & 29.52           & 26.01           & 630k          & -              & 145.89s        \\ \hline

\textbf{MoD-NAS(T=1)} & 33.09          & 30.73          & 27.62          & 31.86          & 29.40          & 26.50          & 32.93          & 30.38          & 26.88          & \textbf{418k}              & \textbf{0.48G}             & \textbf{11.45s} \\ 
\textbf{MoD-NAS(T=3)} &{\ul 33.21}    & {\ul 30.88}    & {\ul 27.83}    & {\ul 31.91}    & {\ul 29.47}    & {\ul 26.59}    & {\ul 33.19}    & {\ul 30.73}    & 27.37       & 1253k      & 1.45G             & {\ul 18.12s} \\
\textbf{MoD-NAS(T=6)} & \textbf{33.28} & \textbf{30.95} & \textbf{27.87} & \textbf{31.94} & \textbf{29.50} & \textbf{26.61} & \textbf{33.34} & \textbf{30.88} & \textbf{27.54} & 2506k             & 2.91G             & 33.39s\\ \hline
\end{tabular}
\end{minipage}}}
\vspace{-0.2in}
\end{table*}
\noindent \textbf{Benefits of Searching for Network Depth.}
 We also have conducted experiments to verify the validity of proposed densely connected block for searching the depth of network. As shown in Fig.~\ref{fig:benifit_depth}, the blue line shows the search with dense connected search strategies. It can be seen that the search procedure is more stable than adding skips (orange line) and the derived network converges rapidly (while adding skips in candidates converges to a shallow network).

\vspace{-3mm}
\subsection{Comparisons with State-of-the-art Methods}
\vspace{-1mm}
For image denoising, we have compared MoD-NAS with several current state-of-the-art methods on three commonly used datasets. The average PSNR results of the benchmark methods in Tab.~\ref{tab:guassian_denoising} are either directly cited from the original papers or reproduced by running the officially released source codes. The testing time is shown in Tab.~\ref{tab:guassian_denoising} is the total time of evaluating the whole Urban100 dataset when $\sigma=25$. To gain deeper insight into the deep unfolding framework, we have considered three different $T$ values and compared their PSNR performance. We have shown the experimental results of $T=1$, $T=3$ and $T=6$ in Tab.~\ref{tab:guassian_denoising}.

\begin{figure}[htbh]

	\newlength\fsfourteen
	\setlength{\fsfourteen}{-4.7mm}
	\scriptsize
	\centering
	\begin{tabular}{cc}
	\tiny
		\hspace{-0.4cm}
		\begin{adjustbox}{valign=t}
			\begin{tabular}{cccccccc}
			    \includegraphics[width=0.083\textwidth]{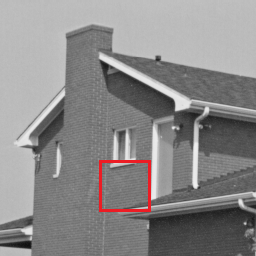} \hspace{\fsfourteen} &
				\includegraphics[width=0.083\textwidth]{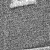} \hspace{\fsfourteen} &
				\includegraphics[width=0.083\textwidth]{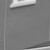}  \hspace{\fsfourteen} &
				\includegraphics[width=0.083\textwidth]{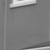} \hspace{\fsfourteen} &
				\includegraphics[width=0.083\textwidth]{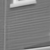} \hspace{\fsfourteen} &
				\includegraphics[width=0.083\textwidth]{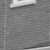} \hspace{\fsfourteen} &
				\\
				
				\includegraphics[width=0.083\textwidth]{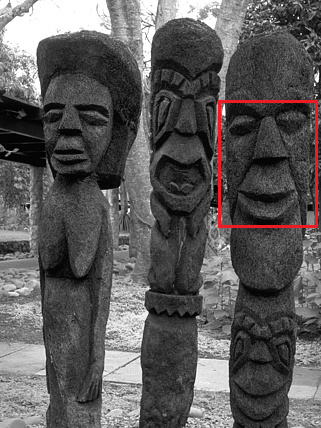} \hspace{\fsfourteen} &
				\includegraphics[width=0.083\textwidth]{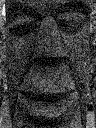} \hspace{\fsfourteen} &
				\includegraphics[width=0.083\textwidth]{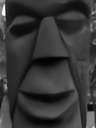}  \hspace{\fsfourteen} &
				\includegraphics[width=0.083\textwidth]{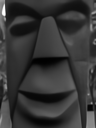} \hspace{\fsfourteen} &
				\includegraphics[width=0.083\textwidth]{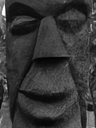} \hspace{\fsfourteen} &
				\includegraphics[width=0.083\textwidth]{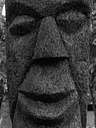} \hspace{\fsfourteen} &
				\\
				\includegraphics[width=0.083\textwidth]{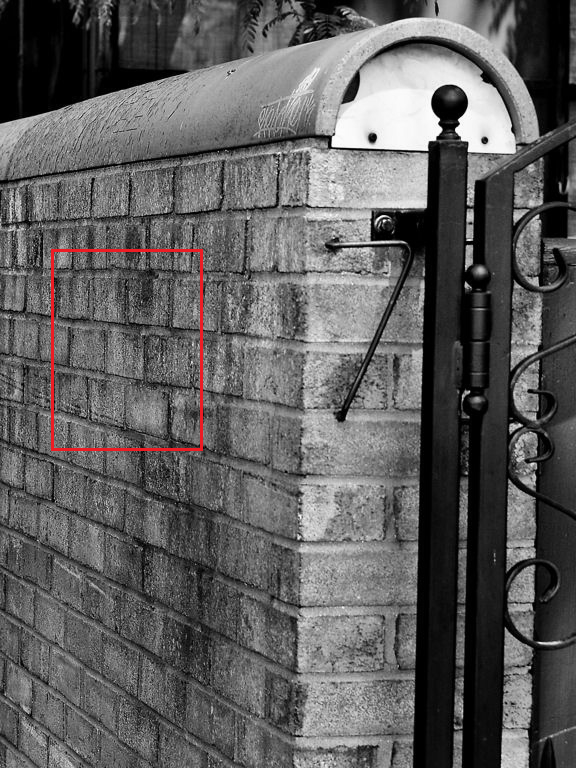} \hspace{\fsfourteen} &
				\includegraphics[width=0.083\textwidth]{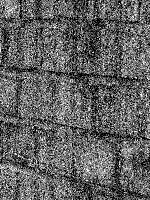} \hspace{\fsfourteen} &
				\includegraphics[width=0.083\textwidth]{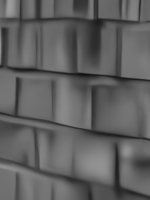}  \hspace{\fsfourteen} &
				\includegraphics[width=0.083\textwidth]{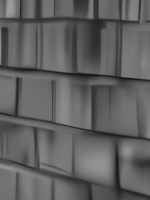} \hspace{\fsfourteen} &
				\includegraphics[width=0.083\textwidth]{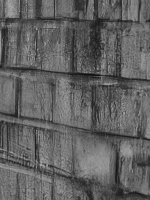} \hspace{\fsfourteen} &
				\includegraphics[width=0.083\textwidth]{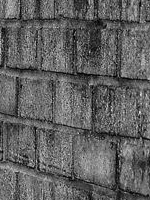} \hspace{\fsfourteen} &
				\\
				GT \hspace{\fsfourteen} &
				Noisy \hspace{\fsfourteen} &
				DPDNN~\cite{dong:TPAMI:2018DPDNN} \hspace{\fsfourteen} &
				RDN~\cite{zhang2020residual} \hspace{\fsfourteen} &
				\textbf{MoD-NAS(T=3)} \hspace{\fsfourteen} &
				GT \hspace{\fsfourteen} &
				\\

				\\
			\end{tabular}
		\end{adjustbox}

	\end{tabular}
	\vspace{-3mm}
	\caption{
		Denoising visual quality comparison. The first row shows the comparison for an image from Set12 with $\sigma=15$; the second row shows the comparison for an image from BSD68 with $\sigma=25$; the third row shows the comparison Urban100 for an image from with $\sigma=50$ (zoom in for better view).
	}
\vspace{-3mm}
	\label{fig:guassian_denoising}
\end{figure}

It can be observed that our MoD-NAS is consistently superior to other competing methods in terms of PSNR performance for all datasets. More importantly, our single-stage model (MoD-NAS(T=1)) has achieved comparable and even better performance than most benchmark methods with the fewest parameters, the lowest number of flops, and the least amount of testing time. As for bigger models, MoD-NAS(T=3) and MoD-NAS(T=6) have achieved better results than other benchmark methods with a larger margin as shown in Tab.~\ref{tab:guassian_denoising}. Although the number of parameters in MoD-NAS(T=3) and MoD-NAS(T=6) is comparable to those in previous methods such as N3Net \cite{plotz2018neural}, DPDNN\cite{dong:TPAMI:2018DPDNN}, MoD-NAS(T=3) and MoD-NAS(T=6) have a great advantage over others in terms of flops and testing time. For instance, our finalized MoD-NAS(T=3) has $1253k$ parameters, which is only
$\textbf{5.7\%}$ that of RDN \cite{zhang2020residual} and $\textbf{43\%}$ that of MemNet\cite{tai2017memnet}. When compared with RDN \cite{zhang2020residual}, our MoD-NAS(T=3) reduces the testing time on Urban100 dataset by as much as $\textbf{95.3\%}$ with even better PSNR results. When compared with model-guided method DPDNN \cite{dong:TPAMI:2018DPDNN} that the denoising network was manually designed based on U-net, our single-stage network MoD-NAS(T=1) has gained $0.09dB$ over DPDNN \cite{dong:TPAMI:2018DPDNN} on the average with only $\textbf{30.7\%}$ parameters, 
$\textbf{9.1\%}$ flops and $\textbf{16.7\%}$ testing time. We have also shown the visual comparison of denoised images with competing methods in Fig.~\ref{fig:guassian_denoising}, from which one can clearly observe the superiority of our method in terms of more fine-detailed texture patterns. The average SSIM and more visual results can be found in appendix. 

\vspace{-3mm}

\subsection{Comparisons with Other NAS Methods}
\begin{table}[htbh]
\centering{
\caption{Comparison of other NAS methods on BSD200 dataset with $\sigma=50$.
 }
\label{tab:other_NAS}
\resizebox{1.0\columnwidth}{!}{
\begin{tabular}{|c|c|c|c|c|c|c|c|c|}
\hline
Models  &PSNR &GPU (Memory) &search+training time  &search strategy\\ \hline
  E-CAE  &26.17 &4 Tesla P100 (16GB) &96 hours &EA\\
  HiNAS  &26.77 &1 Tesla V100 (32GB) &16.5 hours &gradient\\
  \textbf{MoD-NAS(T=3)}  &27.26 &1 RTX 2080Ti (11GB) &17 hours  &gradient\\
  \hline
\end{tabular}}}
\end{table}
Only a small number of NAS methods (e.g., image super-resolution \cite{guo2020hierarchical,chu2019fast} and denoising  \cite{suganuma2018exploiting,zhang2020memory}) have been proposed for low-level vision tasks. Here we compare our proposed MoD-NAS with E-NAS \cite{suganuma2018exploiting} and HiNAS \cite{zhang2020memory} in Tab.~\ref{tab:guassian_denoising} and Tab.~\ref{tab:other_NAS}. Apparently, in Tab.~\ref{tab:other_NAS}, methods with search strategy based on gradient descent have advantages on the cost of GPU memory and searching/training time. When compared with HiNAS~\cite{zhang2020memory}, our proposed method has the following main advantages.
\begin{itemize}
 \vspace{-3mm}
  \item We have proposed a new efficient search space under model-guided framework, which deep denoiser is based on U-net to address the problem that networks found by cell-based search spaces which HiNAS adopted often suffer from long testing time. As shown in Tab.~\ref{tab:guassian_denoising}, MoD-NAS(T=1) takes $\textbf{7.8\%}$ testing time of HiNAS and haves less parameters than HiNAS, which demonstrates MoD-NAS achieves lightweight and low inferring time simultaneously with more competent performance than HiNAS. 
  \vspace{-3mm} 
  \item Compared with HiNAS and E-CAE in Tab.~\ref{tab:guassian_denoising} and Tab.~\ref{tab:other_NAS},  our searched networks MoD-NAS(T=1,3,6) achieve much better performance in terms of PSNR, which indicates the superiority of MoD-NAS
  \vspace{-3mm}
  \item By employing a new highly reusable width search strategy for searching the network width, our supernet can be searched by using one $11GB$ 2080Ti GPU while HiNAS~\cite{zhang2020memory} is trained by using one $32GB$ V100 GPU. 
\end{itemize}

\subsection{Real image denoising on SIDD Dataset.}
\begin{figure}[htbh]

	\newlength\sidd
	\setlength{\sidd}{-4.7mm}
	\scriptsize
	\centering
	\begin{tabular}{cc}
	\tiny
		\hspace{-0.4cm}
		\begin{adjustbox}{valign=t}
			\begin{tabular}{ccccccc}
				\includegraphics[width=0.1\textwidth]{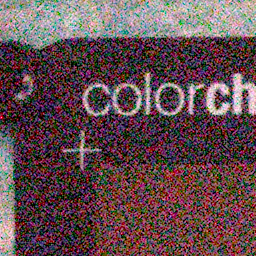} \hspace{\sidd} &
				\includegraphics[width=0.1\textwidth]{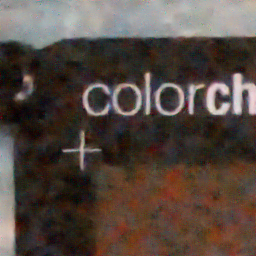}  \hspace{\sidd} &
				\includegraphics[width=0.1\textwidth]{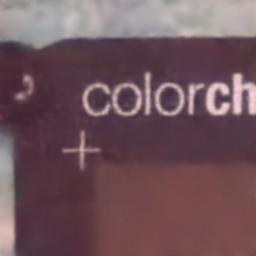} \hspace{\sidd} &
				\includegraphics[width=0.1\textwidth]{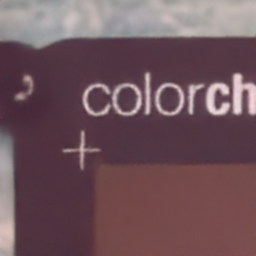} \hspace{\sidd} &
				\includegraphics[width=0.1\textwidth]{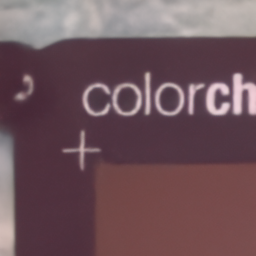} \hspace{\sidd} &
				\\
			    \includegraphics[width=0.1\textwidth]{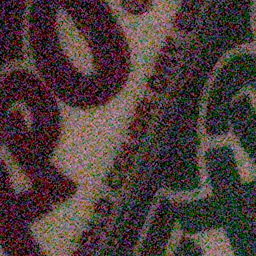} \hspace{\sidd} &
				\includegraphics[width=0.1\textwidth]{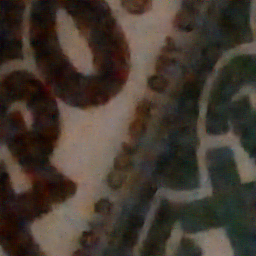}  \hspace{\sidd} &
				\includegraphics[width=0.1\textwidth]{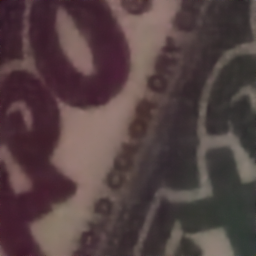} \hspace{\sidd} &
				\includegraphics[width=0.1\textwidth]{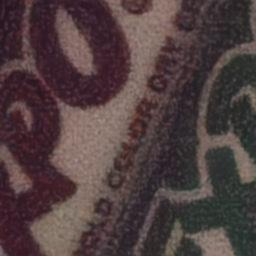} \hspace{\sidd} &
				\includegraphics[width=0.1\textwidth]{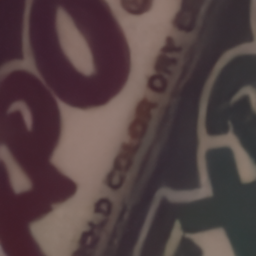} \hspace{\sidd} &
				\\
				Noisy \hspace{\sidd} &
				CBDNet~\cite{guo2019toward} \hspace{\sidd} &
				VDN~\cite{yue2019variational} \hspace{\sidd} &
				DANet~\cite{yue2020dual} \hspace{\sidd} &
				\textbf{MoD-NAS(T=3)} \hspace{\sidd} &
				\\
			\end{tabular}
		\end{adjustbox}

	\end{tabular}
	\caption{Real image denoising visual quality comparison on SIDD testing set(zoom in for better view).
	}
	\label{fig:sidd_denoising}
\end{figure}

To demonstrate the generalization ability of our searched network, we evaluate the performance of MoD-NAS(T=3) on a real blind denoising task with SIDD \cite{abdelhamed2018a} benchmark. SIDD dataset has provided one medium training set (320 image pairs) and a validation set (40 image pairs) for fast training and evaluation, but the testing results can only be obtained by online submission. We have trained MoD-NAS(T=3) on the medium training set and obtained results of testing sets by online submission. Tab.~\ref{tab:SIDD} shows the PSNR and SSIM results of different methods on SIDD testing set. 
Note that the results on testing sets are cited from the official website\footnote{
https://www.eecs.yorku.ca/~kamel/sidd/benchmark.php}. From Tab.~\ref{tab:SIDD}, we can see that MoD-NAS(T=3) can achieve promising results in terms of PSNR and SSIM comparing with other methods. We have shown the real image denoising visual comparison on SIDD dataset in Fig.~\ref{fig:sidd_denoising}, from which we can see that our method has achieved a better result than other methods (e.g., more effective noise suppression). 
\begin{table}[]
\centering{
\caption{Comparison of different methods on SIDD testing set.
 }
\label{tab:SIDD}
\resizebox{1.0\columnwidth}{!}{
\begin{tabular}{|c|c|c|c|c|c|c|}
\hline
Methods & CBDNet \cite{guo2019toward}  & VDN \cite{yue2019variational} & DANet \cite{yue2020dual} & \textbf{MoD-NAS(T=3)} \\ \hline
PSNR    & 33.28  &39.26 &39.25 &39.29              \\ \hline
SSIM    & 0.868  &0.955 &0.955 &0.955              \\ \hline
params  & 4346k  &2325k &9154k &1253k              \\ \hline
\end{tabular}}}
\end{table}
\vspace{-3mm}

\subsection{Image Compression Artifact Reduction.}
We apply the proposed MoD-NAS search method to image compression artifact reduction for further evaluating the generalization capability of our method. We employ the same searching and training setting as the denoising experiments. The results of JPEG quality $q=20$ are listed in Tab.~\ref{tab:Artifact Reduction} and shown in Fig.~\ref{fig:Artifact Reduction}. As shown in Tab.~\ref{tab:Artifact Reduction}, the MoD-NAS-AR searched by MoD-NAS achieves much better performance than other methods. When compared with RDN, MoD-NAS-AR gains $\textbf{0.23}dB$ improvement in terms of PSNR with only $\textbf{7.6\%}$ parameters of RDN. More visual comparisons can be found in appendix.
\begin{table}[]
\caption{Comparisons of different methods on image compression artifact reduction with $q=20$ on LIVE1 dataset.
 }
\resizebox{1.0\columnwidth}{!}{
\begin{tabular}{|c|c|c|c|c|c|}
\hline
Methods & TNRD \cite{chen:TPAMI:2017TNRD}   & DnCNN \cite{zhang2017beyond}  & MemNet \cite{tai2017memnet} & RDN \cite{zhang2020residual}    & \textbf{MoD-NAS-AR} \\ \hline
PSNR    & 31.46  & 31.59  & 31.83  & 32.07  & 32.30               \\ \hline
SSIM    & 0.8769 & 0.8802 & 0.8846 & 0.8882 & 0.8945              \\ \hline
params  & -      & 668k   & 2905k  & 21970k & 1670K               \\ \hline
\end{tabular}}
\label{tab:Artifact Reduction}
\end{table}
\vspace{-3mm}

\begin{figure}[htbh]

	\newlength\deblocking
	\setlength{\deblocking}{-4.7mm}
	\scriptsize
	\centering
	\tiny
		\hspace{-0.7cm}
		\begin{adjustbox}{valign=t}
			\begin{tabular}{ccccccc}
				\includegraphics[width=0.099\textwidth]{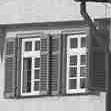}  \hspace{\deblocking} &
				\includegraphics[width=0.099\textwidth]{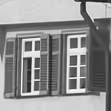} \hspace{\deblocking} &
				\includegraphics[width=0.099\textwidth]{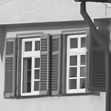} \hspace{\deblocking} &
				\includegraphics[width=0.099\textwidth]{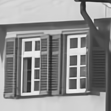} \hspace{\deblocking} &
				\includegraphics[width=0.099\textwidth]{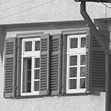} \hspace{\deblocking} &
				\\
				JEPG(q=20) \hspace{\deblocking} &
				MemNet~\cite{tai2017memnet} \hspace{\deblocking} &
				RDN~\cite{zhang:CVPR:2018RDN} \hspace{\deblocking} &
				\textbf{MoD-NAS-AR} \hspace{\deblocking} &
				GT ~ \hspace{\deblocking} &
				\\
			\end{tabular}
		\end{adjustbox}

	\caption{Image compression Artifact reduction visual quality comparison on LIVE1 dataset (zoom in for better view).
	}
	\label{fig:Artifact Reduction}
\end{figure}

\vspace{-0.2in}
\section{Conclusion}
In this paper, we have presented a novel MoD-NAS based approach to image denoising. By incorporating the strengths of model-guided design and NAS, we have constructed a new search space and designed flexible search strategies specially tailored for the task of image denoising. Through searching in the space of concatenated U-net, we demonstrate how the joint consideration of layer operation, network width, and network depth can lead to a network solution with excellent performance including visual quality and convergence property.  
Our proposed network could achieve at least comparable and often even better PSNR results than current leading methods with less number of parameters and flops as well as less amount of testing time.
{\small
\bibliographystyle{ieee_fullname}
\bibliography{ning}

\begin{thebibliography}{10}\itemsep=-1pt

\bibitem{abdelhamed2018a}
Abdelrahman {Abdelhamed}, Stephen {Lin}, and Michael~S. {Brown}.
\newblock A high-quality denoising dataset for smartphone cameras.
\newblock In {\em 2018 IEEE/CVF Conference on Computer Vision and Pattern
  Recognition}, pages 1692--1700, 2018.

\bibitem{cai2018proxylessnas}
Han {Cai}, Ligeng {Zhu}, and Song {Han}.
\newblock Proxylessnas: Direct neural architecture search on target task and
  hardware.
\newblock In {\em International Conference on Learning Representations}, 2018.

\bibitem{chen:TPAMI:2017TNRD}
Yunjin Chen and Thomas Pock.
\newblock Trainable nonlinear reaction diffusion: A flexible framework for fast
  and effective image restoration.
\newblock {\em IEEE Transactions on Pattern Analysis Machine Intelligence},
  39(6):1256--1272, 2017.

\bibitem{chen2019detnas}
Yukang {Chen}, Tong {Yang}, Xiangyu {Zhang}, Gaofeng {Meng}, Chunhong {Pan},
  and Jian {Sun}.
\newblock Detnas: Neural architecture search on object detection.
\newblock 2019.

\bibitem{chu2019fast}
Xiangxiang Chu, Bo Zhang, Hailong Ma, Ruijun Xu, Jixiang Li, and Qingyuan Li.
\newblock Fast, accurate and lightweight super-resolution with neural
  architecture search.
\newblock {\em arXiv preprint arXiv:1901.07261}, 2019.

\bibitem{dabov:TIP:2007BM3D}
Kostadin Dabov, Alessandro Foi, Vladimir Katkovnik, and Karen Egiazarian.
\newblock Image denoising by sparse 3-d transform-domain collaborative
  filtering.
\newblock {\em IEEE Transactions on image processing}, 16(8):2080--2095, 2007.

\bibitem{dong2020multi}
Hang {Dong}, Jinshan {Pan}, Lei {Xiang}, Zhe {Hu}, Xinyi {Zhang}, Fei {Wang},
  and Ming-Hsuan {Yang}.
\newblock Multi-scale boosted dehazing network with dense feature fusion.
\newblock In {\em 2020 IEEE/CVF Conference on Computer Vision and Pattern
  Recognition (CVPR)}, pages 2157--2167, 2020.

\bibitem{dong:TPAMI:2018DPDNN}
Weisheng Dong, Peiyao Wang, Wotao Yin, Guangming Shi, Fangfang Wu, and Xiaotong
  Lu.
\newblock Denoising prior driven deep neural network for image restoration.
\newblock {\em IEEE transactions on pattern analysis and machine intelligence},
  41(10):2305--2318, 2018.

\bibitem{dong2013sparse}
Weisheng Dong, Lei Zhang, Rastislav Lukac, and Guangming Shi.
\newblock Sparse representation based image interpolation with nonlocal
  autoregressive modeling.
\newblock {\em IEEE Transactions on Image Processing}, 22(4):1382--1394, 2013.

\bibitem{fang2020densely}
Jiemin Fang, Yuzhu Sun, Qian Zhang, Yuan Li, Wenyu Liu, and Xinggang Wang.
\newblock Densely connected search space for more flexible neural architecture
  search.
\newblock In {\em Proceedings of the IEEE/CVF Conference on Computer Vision and
  Pattern Recognition}, pages 10628--10637, 2020.

\bibitem{guo2019toward}
Shi {Guo}, Zifei {Yan}, Kai {Zhang}, Wangmeng {Zuo}, and Lei {Zhang}.
\newblock Toward convolutional blind denoising of real photographs.
\newblock In {\em 2019 IEEE/CVF Conference on Computer Vision and Pattern
  Recognition (CVPR)}, pages 1712--1722, 2019.

\bibitem{guo2020hierarchical}
Yong {Guo}, Yongsheng {Luo}, Zhenhao {He}, Jin {Huang}, and Jian {Chen}.
\newblock Hierarchical neural architecture search for single image
  super-resolution.
\newblock {\em IEEE Signal Processing Letters}, 27:1255--1259, 2020.

\bibitem{jia2019focnet}
Xixi {Jia}, Sanyang {Liu}, Xiangchu {Feng}, and Lei {Zhang}.
\newblock Focnet: A fractional optimal control network for image denoising.
\newblock In {\em 2019 IEEE/CVF Conference on Computer Vision and Pattern
  Recognition (CVPR)}, pages 6054--6063, 2019.

\bibitem{kim:TPAMI:2010sparse}
Kwang~In {Kim} and Younghee {Kwon}.
\newblock Single-image super-resolution using sparse regression and natural
  image prior.
\newblock {\em IEEE Transactions on Pattern Analysis and Machine Intelligence},
  32(6):1127--1133, 2010.

\bibitem{kingma2014adam}
Diederik~P Kingma and Jimmy Ba.
\newblock Adam: A method for stochastic optimization.
\newblock {\em arXiv preprint arXiv:1412.6980}, 2014.

\bibitem{liang2020darts+}
Hanwen Liang, Shifeng Zhang, Jiacheng Sun, Xingqiu He, Weiran Huang, Kechen
  Zhuang, and Zhenguo Li.
\newblock Darts+: Improved differentiable architecture search with early
  stopping, 2020.

\bibitem{lim:CVPR:2017:EDSR}
Bee {Lim}, Sanghyun {Son}, Heewon {Kim}, Seungjun {Nah}, and Kyoung~Mu {Lee}.
\newblock Enhanced deep residual networks for single image super-resolution.
\newblock In {\em 2017 IEEE Conference on Computer Vision and Pattern
  Recognition Workshops (CVPRW)}, pages 1132--1140, 2017.

\bibitem{liu2019auto}
Chenxi {Liu}, Liang-Chieh {Chen}, Florian {Schroff}, Hartwig {Adam}, Wei {Hua},
  Alan~L. {Yuille}, and Li {Fei-Fei}.
\newblock Auto-deeplab: Hierarchical neural architecture search for semantic
  image segmentation.
\newblock In {\em 2019 IEEE/CVF Conference on Computer Vision and Pattern
  Recognition (CVPR)}, pages 82--92, 2019.

\bibitem{liu2018non}
Ding Liu, Bihan Wen, Yuchen Fan, Chen~Change Loy, and Thomas~S Huang.
\newblock Non-local recurrent network for image restoration.
\newblock In {\em Advances in Neural Information Processing Systems}, pages
  1673--1682, 2018.

\bibitem{liu2018hierarchical}
Hanxiao {Liu}, Karen {Simonyan}, Oriol {Vinyals}, Chrisantha {Fernando}, and
  Koray {Kavukcuoglu}.
\newblock Hierarchical representations for efficient architecture search.
\newblock In {\em ICLR 2018 : International Conference on Learning
  Representations 2018}, 2018.

\bibitem{liu2019darts}
Hanxiao {Liu}, Karen {Simonyan}, and Yiming {Yang}.
\newblock Darts: Differentiable architecture search.
\newblock In {\em ICLR 2019 : 7th International Conference on Learning
  Representations}, 2019.

\bibitem{liu2019dual}
Xing Liu, Masanori Suganuma, Zhun Sun, and Takayuki Okatani.
\newblock Dual residual networks leveraging the potential of paired operations
  for image restoration.
\newblock In {\em Proceedings of the IEEE Conference on Computer Vision and
  Pattern Recognition}, pages 7007--7016, 2019.

\bibitem{loshchilov2016sgdr}
Ilya {Loshchilov} and Frank {Hutter}.
\newblock Sgdr: Stochastic gradient descent with warm restarts.
\newblock In {\em ICLR (Poster)}, 2016.

\bibitem{lu2020beyond}
Xiaotong Lu, Han Huang, Weisheng Dong, Xin Li, and Guangming Shi.
\newblock Beyond network pruning: a joint search-and-training approach.
\newblock In {\em Proc. of IJCAI}.

\bibitem{ma2017waterloo}
Kede Ma, Zhengfang Duanmu, Qingbo Wu, Zhou Wang, Hongwei Yong, Hongliang Li,
  and Lei Zhang.
\newblock {Waterloo Exploration Database}: New challenges for image quality
  assessment models.
\newblock {\em IEEE Transactions on Image Processing}, 26(2):1004--1016, Feb.
  2017.

\bibitem{ning2020accurate}
Q. {Ning}, W. {Dong}, G. {Shi}, L. {Li}, and X. {Li}.
\newblock Accurate and lightweight image super-resolution with model-guided
  deep unfolding network.
\newblock {\em IEEE Journal of Selected Topics in Signal Processing}, pages
  1--1, 2020.

\bibitem{osher2005an}
Stanley~J. {Osher}, Martin {Burger}, Donald {Goldfarb}, Jinjun {Xu}, and Wotao
  {Yin}.
\newblock An iterative regularization method for total variation-based image
  restoration.
\newblock {\em Multiscale Modeling and Simulation}, 4(2):460--489, 2005.

\bibitem{plotz2018neural}
Tobias Pl{\"o}tz and Stefan Roth.
\newblock Neural nearest neighbors networks.
\newblock In {\em Advances in Neural Information Processing Systems}, pages
  1087--1098, 2018.

\bibitem{real2019regularized}
Esteban {Real}, Alok {Aggarwal}, Yanping {Huang}, and Quoc~V. {Le}.
\newblock Regularized evolution for image classifier architecture search.
\newblock {\em Proceedings of the AAAI Conference on Artificial Intelligence},
  33(1):4780--4789, 2019.

\bibitem{sandler2018mobilenetv2}
Mark {Sandler}, Andrew {Howard}, Menglong {Zhu}, Andrey {Zhmoginov}, and
  Liang-Chieh {Chen}.
\newblock Mobilenetv2: Inverted residuals and linear bottlenecks.
\newblock In {\em 2018 IEEE/CVF Conference on Computer Vision and Pattern
  Recognition}, pages 4510--4520, 2018.

\bibitem{sim2019a}
Hyeonjun {Sim} and Munchurl {Kim}.
\newblock A deep motion deblurring network based on per-pixel adaptive kernels
  with residual down-up and up-down modules.
\newblock In {\em 2019 IEEE/CVF Conference on Computer Vision and Pattern
  Recognition Workshops (CVPRW)}, pages 2140--2149, 2019.

\bibitem{suganuma2018exploiting}
Masanori {Suganuma}, Mete {Ozay}, and Takayuki {Okatani}.
\newblock Exploiting the potential of standard convolutional autoencoders for
  image restoration by evolutionary search.
\newblock In {\em ICML 2018: Thirty-fifth International Conference on Machine
  Learning}, pages 4771--4780, 2018.

\bibitem{tai2017memnet}
Ying {Tai}, Jian {Yang}, Xiaoming {Liu}, and Chunyan {Xu}.
\newblock Memnet: A persistent memory network for image restoration.
\newblock In {\em 2017 IEEE International Conference on Computer Vision
  (ICCV)}, pages 4549--4557, 2017.

\bibitem{tan2019mnasnet}
Mingxing {Tan}, Bo {Chen}, Ruoming {Pang}, Vijay {Vasudevan}, Mark {Sandler},
  Andrew {Howard}, and Quoc~V. {Le}.
\newblock Mnasnet: Platform-aware neural architecture search for mobile.
\newblock In {\em 2019 IEEE/CVF Conference on Computer Vision and Pattern
  Recognition (CVPR)}, pages 2820--2828, 2019.

\bibitem{wu2019fbnet}
Bichen {Wu}, Kurt {Keutzer}, Xiaoliang {Dai}, Peizhao {Zhang}, Yanghan {Wang},
  Fei {Sun}, Yiming {Wu}, Yuandong {Tian}, Peter {Vajda}, and Yangqing {Jia}.
\newblock Fbnet: Hardware-aware efficient convnet design via differentiable
  neural architecture search.
\newblock In {\em 2019 IEEE/CVF Conference on Computer Vision and Pattern
  Recognition (CVPR)}, pages 10734--10742, 2019.

\bibitem{xie2020weight}
Lingxi Xie, Xin Chen, Kaifeng Bi, Longhui Wei, Yuhui Xu, Zhengsu Chen, Lanfei
  Wang, An Xiao, Jianlong Chang, Xiaopeng Zhang, et~al.
\newblock Weight-sharing neural architecture search: A battle to shrink the
  optimization gap.
\newblock {\em arXiv preprint arXiv:2008.01475}, 2020.

\bibitem{yang2019frame}
Wenhan {Yang}, Jiaying {Liu}, and Jiashi {Feng}.
\newblock Frame-consistent recurrent video deraining with dual-level flow.
\newblock In {\em 2019 IEEE/CVF Conference on Computer Vision and Pattern
  Recognition (CVPR)}, pages 1661--1670, 2019.

\bibitem{yue2019variational}
Zongsheng {Yue}, Hongwei {Yong}, Qian {Zhao}, Lei {Zhang}, and Deyu {Meng}.
\newblock Variational denoising network: Toward blind noise modeling and
  removal.
\newblock In {\em Advances in Neural Information Processing Systems}, pages
  1690--1701, 2019.

\bibitem{yue2020dual}
Zongsheng {Yue}, Qian {Zhao}, Lei {Zhang}, and Deyu {Meng}.
\newblock Dual adversarial network: Toward real-world noise removal and noise
  generation.
\newblock {\em arXiv: Computer Vision and Pattern Recognition}, 2020.

\bibitem{zhang2020memory}
Haokui {Zhang}, Ying {Li}, Hao {Chen}, and Chunhua {Shen}.
\newblock Memory-efficient hierarchical neural architecture search for image
  denoising.
\newblock In {\em 2020 IEEE/CVF Conference on Computer Vision and Pattern
  Recognition (CVPR)}, pages 3657--3666, 2020.

\bibitem{zhang2020deep}
Kai Zhang, Luc~Van Gool, and Radu Timofte.
\newblock Deep unfolding network for image super-resolution.
\newblock In {\em Proceedings of the IEEE/CVF Conference on Computer Vision and
  Pattern Recognition}, pages 3217--3226, 2020.

\bibitem{zhang2017beyond}
Kai {Zhang}, Wangmeng {Zuo}, Yunjin {Chen}, Deyu {Meng}, and Lei {Zhang}.
\newblock Beyond a gaussian denoiser: Residual learning of deep cnn for image
  denoising.
\newblock {\em IEEE Transactions on Image Processing}, 26(7):3142--3155, 2017.

\bibitem{zhang2017learning}
Kai {Zhang}, Wangmeng {Zuo}, Shuhang {Gu}, and Lei {Zhang}.
\newblock Learning deep cnn denoiser prior for image restoration.
\newblock In {\em 2017 IEEE Conference on Computer Vision and Pattern
  Recognition (CVPR)}, pages 2808--2817, 2017.

\bibitem{zhang2019deep}
Kai Zhang, Wangmeng Zuo, and Lei Zhang.
\newblock Deep plug-and-play super-resolution for arbitrary blur kernels.
\newblock In {\em Proceedings of the IEEE Conference on Computer Vision and
  Pattern Recognition}, pages 1671--1681, 2019.

\bibitem{zhang2018residual}
Yulun Zhang, Yapeng Tian, Yu Kong, Bineng Zhong, and Yun Fu.
\newblock Residual dense network for image super-resolution.
\newblock In {\em Proceedings of the IEEE conference on computer vision and
  pattern recognition}, pages 2472--2481, 2018.

\bibitem{zhang:CVPR:2018RDN}
Yulun Zhang, Yapeng Tian, Yu Kong, Bineng Zhong, and Yun Fu.
\newblock Residual dense network for image super-resolution.
\newblock In {\em Proceedings of the IEEE Conference on Computer Vision and
  Pattern Recognition}, pages 2472--2481, 2018.

\bibitem{zhang2020residual}
Yulun {Zhang}, Yapeng {Tian}, Yu {Kong}, Bineng {Zhong}, and Yun {Fu}.
\newblock Residual dense network for image restoration.
\newblock {\em IEEE Transactions on Pattern Analysis and Machine Intelligence},
  pages 1--1, 2020.

\bibitem{zoph2017neural}
Barret {Zoph} and Quoc {Le}.
\newblock Neural architecture search with reinforcement learning.
\newblock In {\em ICLR 2017 : International Conference on Learning
  Representations 2017}, 2017.

\bibitem{zoph2018learning}
Barret {Zoph}, Vijay {Vasudevan}, Jonathon {Shlens}, and Quoc~V. {Le}.
\newblock Learning transferable architectures for scalable image recognition.
\newblock In {\em 2018 IEEE/CVF Conference on Computer Vision and Pattern
  Recognition}, pages 8697--8710, 2018.

\end{thebibliography}
}
\newpage
 \appendix
\section{More Comparisons of Gaussian image denoising with State-of-the-art Methods}
\begin{figure*}[htbh]

	\newlength\denoising
	\setlength{\denoising}{0mm}
	\scriptsize
	\centering
	\tiny
		\begin{adjustbox}{valign=t}
			\begin{tabular}{cccccccc}
				
				\includegraphics[width=0.22\textwidth]{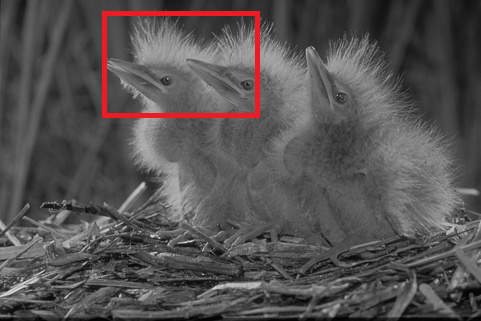} \hspace{\denoising} &
				\includegraphics[width=0.22\textwidth]{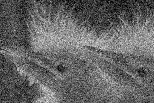}  \hspace{\denoising} &
				\includegraphics[width=0.22\textwidth]{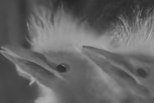} \hspace{\denoising} &
				\includegraphics[width=0.22\textwidth]{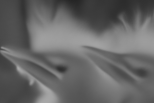} \hspace{\denoising} &
				\\
				`Img\_027' from BSD68 \hspace{\denoising} &
				Noisy with $\sigma=25$\hspace{\denoising} &
				N3Net~\cite{plotz2018neural} \hspace{\denoising} &
				MemNet~\cite{tai2017memnet} \hspace{\denoising} &
				\\
				\includegraphics[width=0.22\textwidth]{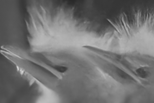} \hspace{\denoising} &
				\includegraphics[width=0.22\textwidth]{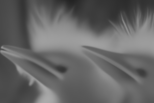} \hspace{\denoising} &
				\includegraphics[width=0.22\textwidth]{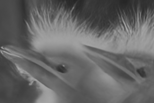} \hspace{\denoising} &
				\includegraphics[width=0.22\textwidth]{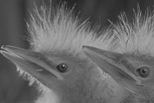} \hspace{\denoising} &
				\\
				DPDNN~\cite{dong:TPAMI:2018DPDNN} \hspace{\denoising} &
				RDN~\cite{zhang:CVPR:2018RDN} \hspace{\denoising} &
				\textbf{MoD-NAS-B} \hspace{\denoising} &
				GT ~ \hspace{\denoising} &
				\\
				\includegraphics[width=0.22\textwidth]{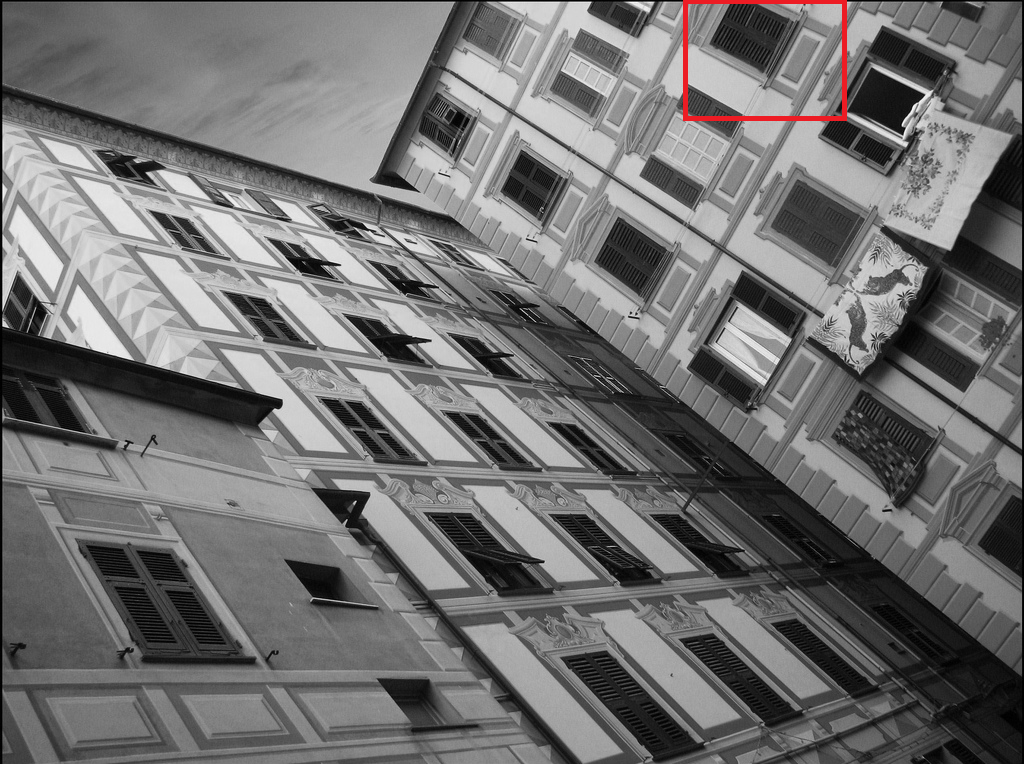} \hspace{\denoising} &
				\includegraphics[width=0.22\textwidth]{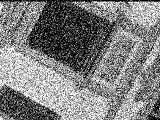}  \hspace{\denoising} &
				\includegraphics[width=0.22\textwidth]{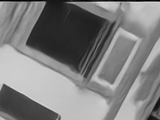} \hspace{\denoising} &
				\includegraphics[width=0.22\textwidth]{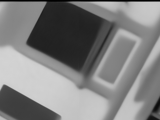} \hspace{\denoising} &
				\\
				`Img\_013' from Urban100 \hspace{\denoising} &
				Noisy with $\sigma=50$\hspace{\denoising} &
				N3Net~\cite{plotz2018neural} \hspace{\denoising} &
				MemNet~\cite{tai2017memnet} \hspace{\denoising} &
				\\
				\includegraphics[width=0.22\textwidth]{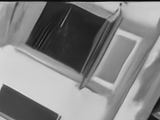} \hspace{\denoising} &
				\includegraphics[width=0.22\textwidth]{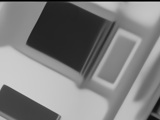} \hspace{\denoising} &
				\includegraphics[width=0.22\textwidth]{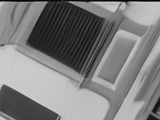} \hspace{\denoising} &
				\includegraphics[width=0.22\textwidth]{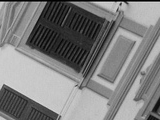} \hspace{\denoising} &
				\\
				DPDNN~\cite{dong:TPAMI:2018DPDNN} \hspace{\denoising} &
				RDN~\cite{zhang:CVPR:2018RDN} \hspace{\denoising} &
				\textbf{MoD-NAS-B} \hspace{\denoising} &
				GT ~ \hspace{\denoising} &
				\\
				\includegraphics[width=0.22\textwidth]{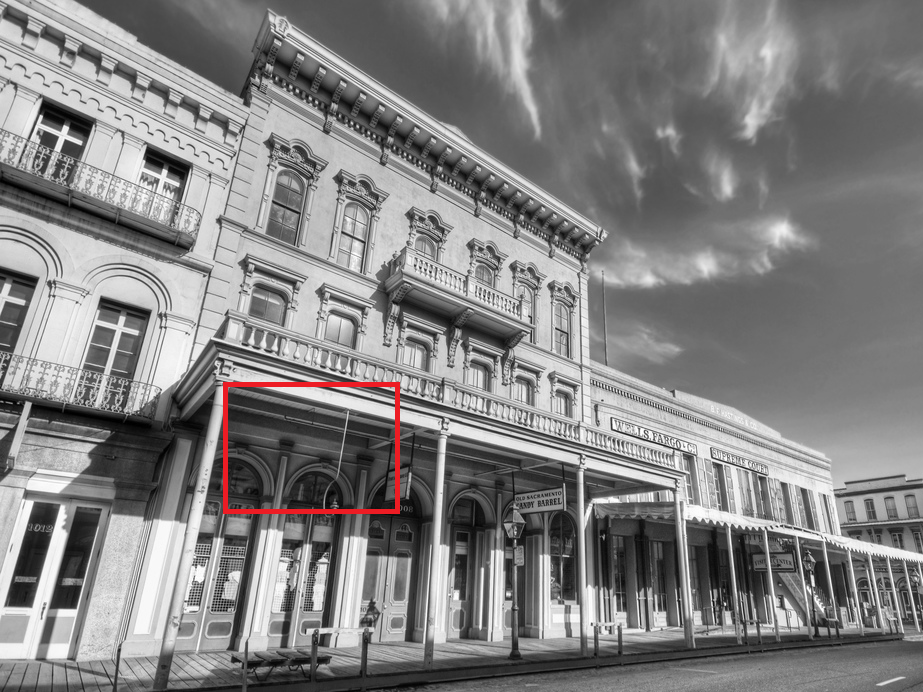} \hspace{\denoising} &
				\includegraphics[width=0.22\textwidth]{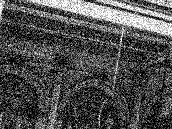}  \hspace{\denoising} &
				\includegraphics[width=0.22\textwidth]{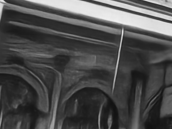} \hspace{\denoising} &
				\includegraphics[width=0.22\textwidth]{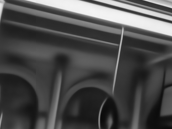} \hspace{\denoising} &
				\\
				`Img\_017' from Urban100 \hspace{\denoising} &
				Noisy with $\sigma=50$\hspace{\denoising} &
				N3Net~\cite{plotz2018neural} \hspace{\denoising} &
				MemNet~\cite{tai2017memnet} \hspace{\denoising} &
				\\
				\includegraphics[width=0.22\textwidth]{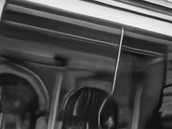} \hspace{\denoising} &
				\includegraphics[width=0.22\textwidth]{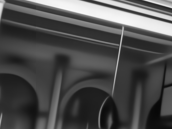} \hspace{\denoising} &
				\includegraphics[width=0.22\textwidth]{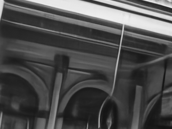} \hspace{\denoising} &
				\includegraphics[width=0.22\textwidth]{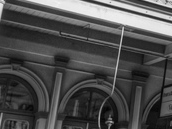} \hspace{\denoising} &
				\\
				DPDNN~\cite{dong:TPAMI:2018DPDNN} \hspace{\denoising} &
				RDN~\cite{zhang:CVPR:2018RDN} \hspace{\denoising} &
				\textbf{MoD-NAS-B} \hspace{\denoising} &
				GT ~ \hspace{\denoising} &
				\\
			\end{tabular}
		\end{adjustbox}

	\caption{Image denoising visual quality comparison (zoom in for better view).
	}
	\label{fig:Gaussian_denoising}
\end{figure*}

Here, we have reported SSIM results and more visual comparisons of Gaussian image denoising results in Tab.~\ref{tab:ssim_guassian_denoising} and Fig.~\ref{fig:Gaussian_denoising} respectively. As shown in Tab.~\ref{tab:ssim_guassian_denoising},  it is easy to see that MoD-NAS is superior to all competing methods in terms of SSIM values. Besides, the visual comparison results with different $\sigma$ are shown in Fig.~\ref{fig:Gaussian_denoising} where our searched network performs better than other competing methods in terms of more recovered image texture details. For example, the first row of Fig.~\ref{fig:Gaussian_denoising} shows the result of `Img\_027’ from BSD68 with $\sigma=25$, where MoD-NAS-B has recovered with fewer visible artifacts and sharper edge of bird hair than other competing methods; the second row of Fig.~\ref{fig:Gaussian_denoising} shows the result of `Img\_013’ from Urban with $\sigma=50$, where our searched network has recovered the stripe shape of wooden window better than others.

\begin{figure}[htbh]
\centering
\includegraphics[width=1.0\linewidth]{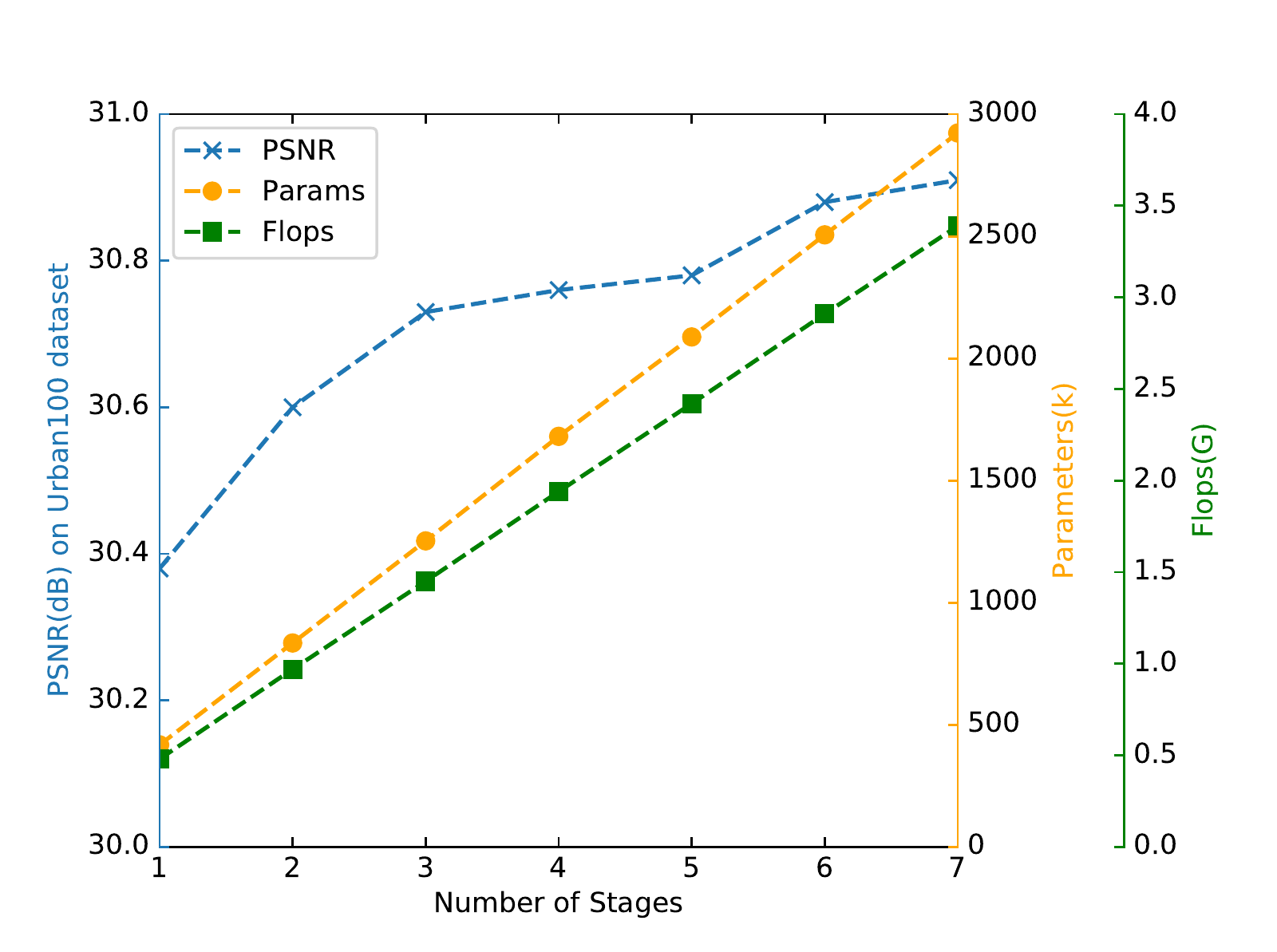}
\caption{The average PSNR performance as a function of parameter $T$ (the total number of Unet stages) of proposed MoG-NAS with $\sigma=25$ on Set12.}
  \label{fig:psnr_stages}
\end{figure}

\begin{table*}[htbh]
\centering{
\caption{Average SSIM results  for \textbf{Gaussian} image denoising on three benchmark datasets.
 The best performance is shown by bold. The testing time is the total time of evaluating the whole Urban100 dataset.}
\label{tab:ssim_guassian_denoising}
\resizebox{2.0\columnwidth}{!}{
 \begin{tabular}{|c|c|c|c|c|c|c|c|c|c|c|c|c|}
\hline
\multirow{2}{*}{Methods} & \multicolumn{3}{c|}{Set12}                      & \multicolumn{3}{c|}{BSD68}                       & \multicolumn{3}{c|}{Urban100}                    & \multirow{2}{*}{params} & \multirow{2}{*}{flops} & \multirow{2}{*}{{\begin{tabular}[c]{@{}c@{}}testing \\ time\end{tabular}}} \\ \cline{2-10}
 
                         & 15                    & 25                    & 50                    & 15                    & 25                    & 50                    & 15                    & 25                    & 50               &                         &                        &                        \\ \hline \hline
 BM3D                    & 0.8952          & 0.8504          & 0.7676          & 0.8717          & 0.8013          & 0.6864          & 0.9220          & 0.8777          & 0.7791     & -                       & -                      &- \\
 TNRD                    & 0.8958          & 0.8512          & 0.7680          & 0.8769          & 0.8093          & 0.6994          & 0.9031          & 0.8473          & 0.7563    & -                       & -                      &-  \\
 DnCNN                   & 0.9031          & 0.8622          & 0.7829          & 0.8907          & 0.8278          & 0.8278          & 0.9255          & 0.8797          & 0.7874    &  556k                    & 1.28G                  &21.45s  \\
 N3Net                   & -                     & 0.8644           & 0.7939                & -                     & 0.7957                & 0.6455                 & -                     & 0.8917                &0.8148      & 706k                    & 1.62G                  &594.38s      \\
 MemNet                  & 0.9001          & 0.8652          & 0.7563          & 0.8848          & 0.7966          & 0.6466          & 0.9264          & 0.8793          & 0.7554   & 2905k                   & 6.69G                  & 93.83s  \\
 DPDNN                   &0.8970                 & 0.8594                & 0.7907                & 0.8738                &0.8123               &  0.7095               &  0.9322               & 0.8937               & 0.8166           & 1363k                   & 5.25G                  &68.62s \\
 RDN                     & 0.9030          & 0.8680          & 0.7504               & 0.8884          & 0.7942          & 0.6431              & 0.9291          & 0.8804          & 0.7657  & 21970k                  & 50.6G                  &386.95s\\ \hline
 \textbf{MoG-NAS(T=1)} & 0.9070          & 0.8689          & 0.7999          & 0.894           & 0.8350          & 0.7325          & 0.9319          & 0.8953          & 0.8163          & \textbf{418k} & \textbf{0.48G} & \textbf{11.45s} \\
\textbf{MoG-NAS(T=3)} & 0.9087          & 0.8720          & 0.8065          & 0.8950          & 0.8365          & 0.7371          & 0.9347          & 0.9011          & 0.8317          & 1253k         & 1.45G          & 18.12s          \\
\textbf{MoG-NAS(T=6)} & \textbf{0.9101} & \textbf{0.8729} & \textbf{0.8070} & \textbf{0.8956} & \textbf{0.8371} & \textbf{0.7376} & \textbf{0.9364} & \textbf{0.9039} & \textbf{0.8358} & 2506k         & 2.91G          & 33.39s                 \\
 \hline
 \end{tabular}}}
\end{table*}

\section{More Visual Comparisons of Image Compress Artifact Reduction with State-of-the-art Methods}
In this section, we have shown more visual comparisons of image compression artifact reduction with $q=20$ in Fig.~\ref{fig:Artifact Reduction}. Taking the second row of Fig.~\ref{fig:Artifact Reduction} as an example, our searched MoD-NAS-AR has recovered the right letters $ITH$ on the helmet with fewer artifacts than other competing methods.
\begin{figure*}[htbh]

	\newlength\deblockingw
	\setlength{\deblockingw}{-1mm}
	\scriptsize
	\centering
	\tiny
		\hspace{-0.4cm}
		\begin{adjustbox}{valign=t}
			\begin{tabular}{ccccccc}
				\includegraphics[width=0.15\textwidth]{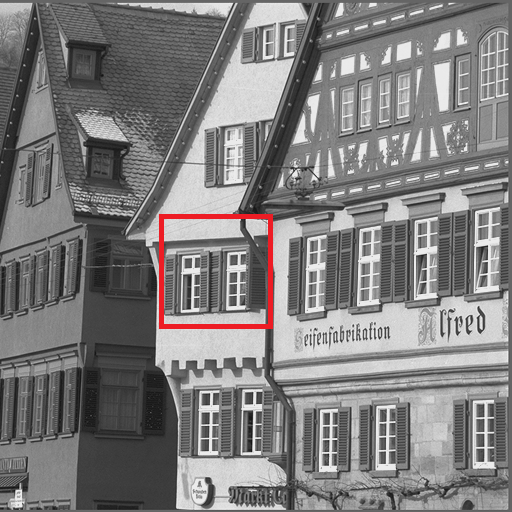} \hspace{\deblockingw} &
				\includegraphics[width=0.15\textwidth]{figs/Noisy_buildings.png}  \hspace{\deblockingw} &
				\includegraphics[width=0.15\textwidth]{figs/MemNet_buildings.png} \hspace{\deblockingw} &
				\includegraphics[width=0.15\textwidth]{figs/RDN_buildings.png} \hspace{\deblockingw} &
				\includegraphics[width=0.15\textwidth]{figs/MoG_NAS_B_buildings.png} \hspace{\deblockingw} &
				\includegraphics[width=0.15\textwidth]{figs/GT_buildings.png} \hspace{\deblockingw} &
				\vspace{3mm}
				\\
				
				\includegraphics[width=0.15\textwidth]{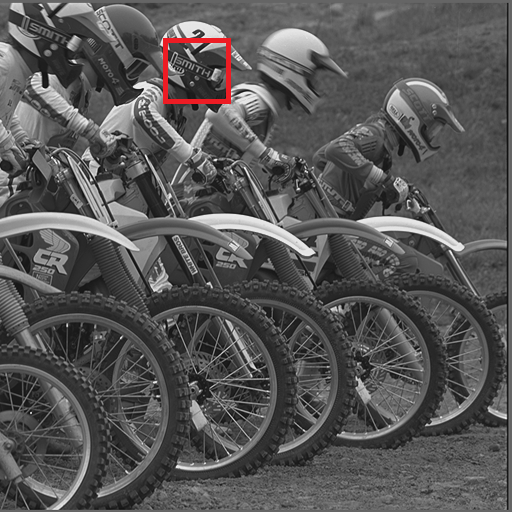} \hspace{\deblockingw} &
				\includegraphics[width=0.15\textwidth]{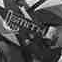}  \hspace{\deblockingw} &
				\includegraphics[width=0.15\textwidth]{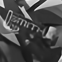} \hspace{\deblockingw} &
				\includegraphics[width=0.15\textwidth]{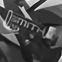} \hspace{\deblockingw} &
				\includegraphics[width=0.15\textwidth]{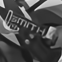} \hspace{\deblockingw} &
				\includegraphics[width=0.15\textwidth]{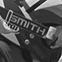} \hspace{\deblockingw} &
				\vspace{3mm}
				\\
				
				\includegraphics[width=0.15\textwidth]{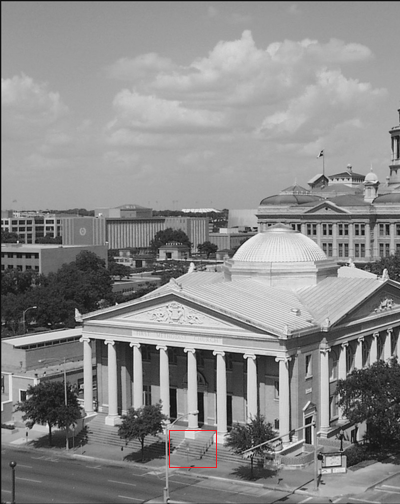} \hspace{\deblockingw} &
				\includegraphics[width=0.15\textwidth]{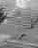}  \hspace{\deblockingw} &
				\includegraphics[width=0.15\textwidth]{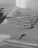} \hspace{\deblockingw} &
				\includegraphics[width=0.15\textwidth]{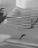} \hspace{\deblockingw} &
				\includegraphics[width=0.15\textwidth]{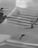} \hspace{\deblockingw} &
				\includegraphics[width=0.15\textwidth]{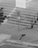} \hspace{\deblockingw} &
				\\
				GT \hspace{\deblockingw} &
				JEPG(q=20) \hspace{\deblockingw} &
				MemNet~\cite{tai2017memnet} \hspace{\deblockingw} &
				RDN~\cite{zhang:CVPR:2018RDN} \hspace{\deblockingw} &
				\textbf{MoD-NAS-AR} \hspace{\deblockingw} &
				GT ~ \hspace{\deblockingw} &
				\\
			\end{tabular}
		\end{adjustbox}

	\caption{Image compression Artifact reduction visual quality comparison on LIVE1 testing set(zoom in for better view).
	}
	\label{fig:Artifact Reduction}
\end{figure*}

\section{Ablation study of impact of stage number T}
To explore the impact of the number of unfolded stages on the denoising performance, we have conducted another experiment with varying the parameter $T$. Fig.~\ref{fig:psnr_stages} shows the average PSNR results of different stages $T$ from two to seven with $\sigma=25$. It can be seen that the PSNR increases as the number of stages increases which indicates that we can choose the suitable number of stages $T$ based on the real application to balance the trade-off between performance and cost.

\section{Details of Searched Architectures}
Architecture details of our searched MoD-NAS-B for image denoising and MoD-NAS-AR for image compression artifact reduction are described in Tab.~\ref{tab:MoD-NAS-B} and Tab.~\ref{tab:MoD-NAS-AR} respectively.
\begin{table*}
  \centering
  \resizebox{0.74\textwidth}{!}{
\begin{tabular}{|c|c|c|c|c|c|c|} \hline
Inputs   size & Operation & kernel\_size & $C_{in}$      & $C_{out}$     & Act & Stride     \\ \hline
$128\times128\times1$     & Conv      & $3\times3$          & 1          & 48         & Relu         & 1          \\
$128\times128\times48$    & Conv      & $3\times3$          & 48         & 48         & Relu         & 1          \\ \hline
$128\times128\times48$    & Conv      & $3\times3$          & 48         & 40         & Relu         & 1          \\
$128\times128\times40$    & Residual block        & $3\times3$          & 40         & 40         & Relu         & 1          \\
$128\times128\times40$    & Residual block        & $3\times3$          & 40         & 40         & Relu         & 1          \\
$128\times128\times40$    & Conv      & $3\times3$          & 40         & 24         & Relu         & 2          \\ \hline
$64\times64\times24$      & Conv      & $3\times3$          & 24         & 32         & Relu         & 1          \\
$64\times64\times32$      & Separable Conv        & $5\times5$          & 32         & 40         & Relu         & 1          \\
$64\times64\times40$      & skip      & -            & -          & -          & -            & - \\
$64\times64\times40$      & Nearest interpolation        & -            & 40         & 32         & -            & -          \\ \hline
$32\times32\times32$      & Conv      & $5\times5$          & 32         & 40         & Relu         & 1          \\
$32\times32\times40$      & skip      & -            & -          & -          & -            & -          \\
$32\times32\times40$      & Residual block        & $3\times3$          & 40         & 40         & Relu         & 1          \\
$32\times32\times40$      & Nearest interpolation        & -            & 40         & 32         & -            & -          \\ \hline
$16\times16\times32$      & Conv      & $5\times5$          & 32         & 48         & Relu         & 1          \\
$16\times16\times48$      & Dilated Conv        & $3\times3$          & 48         & 48         & Relu         & 1          \\
$16\times16\times48$      & skip      & -            & -          & -          & -            & -          \\
$16\times16\times48$      & Bilinear interpolation        & -            & 48         & 40         & -            & -          \\ \hline
$32\times32\times72$      & Conv      & $3\times3$           & 72          & 36         & -            & 1          \\
$32\times32\times36$      & skip      & -            & -          & -          & -            & -          \\
$32\times32\times36$      & Dilated Conv        & $5\times5$          & 36         & 40         & Relu         & 1      \\
$32\times32\times40$      & skip      & -            & -          & -          & -            & -          \\
$32\times32\times40$      & Bilinear interpolation        & -            & 40         & 40         & -            & -          \\ \hline
$64\times64\times72$      & Conv      & $3\times3$           & 72          & 36         & -            & 1          \\
$64\times64\times36$      & Residual block        & $3\times3$          & 36         & 32         & Relu         & 1          \\
$64\times64\times32$      & skip       & -            & -          & -          & -            & - \\
$64\times64\times32$      & Residual block        & $3\times3$          & 32         & 32         & Relu         & 1          \\
$64\times64\times32$      & Bilinear interpolation        & -            & 32         & 32         & -            & -          \\ \hline
$128\times128\times56$      & Conv      & $3\times3$           & 56          & 28         & -            & 1          \\
$128\times128\times28$    & Residual block        & $3\times3$          & 28         & 40         & Relu         & 1          \\
$128\times128\times40$    & Residual block        & $3\times3$          & 40         & 40         & Relu         & 1          \\
$128\times128\times40$    & skip      & -            & -          & -          & -            & - \\ \hline
$128\times128\times40$    & Conv      & $1\times1$          & 40         & 1          & Relu         & 1 \\\hline
\end{tabular}}
  \vspace{0.1cm}
  \caption{Architecture details of MoD-NAS-B for image denoising.}
  \label{tab:MoD-NAS-B}
\end{table*}

\begin{table*}
  \centering
  \resizebox{0.74\textwidth}{!}{
\begin{tabular}{|c|c|c|c|c|c|c|} \hline
Inputs   size & Operation & kernel\_size & $C_{in}$      & $C_{out}$     & Act & Stride     \\ \hline
$128\times128\times1$     & Conv      & $3\times3$          & 1          & 48         & Relu         & 1          \\
$128\times128\times48$    & Conv      & $3\times3$          & 48         & 48         & Relu         & 1          \\ \hline
$128\times128\times48$    & Residual block        & $3\times3$          & 48         & 48         & Relu         & 1           \\
$128\times128\times48$    & skip      & -            & -          & -          & -            & -          \\
$128\times128\times48$    & Residual block        & $3\times3$          & 48         & 40         & Relu         & 1          \\
$128\times128\times40$    & Conv      & $3\times3$          & 40         & 48         & Relu         & 2          \\ \hline
$64\times64\times48$      & Dilated Conv        & $5\times5$          & 48         & 32         & Relu         & 1           \\
$64\times64\times32$      & skip      & -            & -          & -          & -            & -         \\
$64\times64\times32$      & Separable Conv        & $5\times5$          & 32         & 48         & Relu         & 1  \\
$64\times64\times48$      & Area interpolation        & -            & 48         & 48         & -            & -          \\ \hline
$32\times32\times48$      & skip      & -            & -          & -          & -            & -          \\
$32\times32\times48$      & Residual block        & $3\times3$          & 48         & 40         & Relu         & 1          \\
$32\times32\times40$      & skip      & -            & -          & -          & -            & -                 \\
$32\times32\times40$      & Conv      & $3\times3$          & 40         & 48         & Relu         & 2         \\ \hline
$16\times16\times48$      & Separable Conv        & $5\times5$          & 48         & 48         & Relu         & 1          \\
$16\times16\times48$      & Conv        & $5\times5$          & 48         & 32         & Relu         & 1          \\
$16\times16\times32$      & skip      & -            & -          & -          & -            & -          \\
$16\times16\times32$      & Bilinear interpolation        & -            & 32         & 48         & -            & -          \\ \hline
$32\times32\times96$      & Conv      & $3\times3$           & 96          & 48         & -            & 1          \\
$32\times32\times48$      & Residual block        & $3\times3$          & 48         & 48         & Relu         & 1         \\
$32\times32\times48$      & Dilated Conv        & $5\times5$          & 48         & 48       & Relu         & 1          \\
$32\times32\times48$      & Conv        & $5\times5$          & 48         & 32        & Relu         & 1           \\
$32\times32\times32$      & Nearest interpolation        & -            & 32         & 48         & -            & -          \\ \hline
$64\times64\times96$      & Conv      & $3\times3$           & 96          & 48         & -            & 1          \\
$64\times64\times48$      & skip       & -            & -          & -          & -            & -          \\
$64\times64\times48$      & Conv        & $3\times3$          & 48         & 48         & Relu         & 1\\
$64\times64\times32$      & Residual block        & $3\times3$          & 32         & 48         & Relu         & 1          \\
$64\times64\times48$      & Deconvolution        & $3\times3$         & 48         & 40         & Relu            & 2          \\ \hline
$128\times128\times88$      & Conv      & $3\times3$           & 88          & 44        & -            & 1          \\
$128\times128\times44$    & Dilated Conv        & $3\times3$          & 44         & 40         & Relu         & 1          \\
$128\times128\times40$    & Residual block        & $3\times3$          & 40         & 40         & Relu         & 1          \\
$128\times128\times40$    & Conv        & $3\times3$          & 40         & 40         & Relu         & 1 \\ \hline
$128\times128\times40$    & Conv      & $1\times1$          & 40         & 1          & Relu         & 1 \\ \hline
\end{tabular}}
  \vspace{0.1cm}
  \caption{Architecture details of MoD-NAS-AR for compression artifact reduction.}
  \label{tab:MoD-NAS-AR}
\end{table*}

\end{document}